\documentclass{fundam}

\usepackage{url} %
\usepackage[ruled,lined]{algorithm2e}%
\usepackage{graphicx}%
\usepackage{amsmath,amssymb,xypic,enumitem}
\usepackage{tikz}
\usetikzlibrary{automata, positioning, arrows}
\input{gtaba.mac}

\begin{document}

\setcounter{page}{115}
\publyear{22}
\papernumber{2106}
\volume{185}
\issue{2}

  \finalVersionForIOS

\title{Getting There and Back Again}

\author{Olivier Danvy\thanks{Address for correspondence: Yale-NUS College \& School of Computing,
                                       National University of Singapore. \newline \newline
          \vspace*{-6mm}{\scriptsize{Received August 2020; \ accepted April 2022.}}}
\\
  Yale-NUS College \& School of Computing\\ %
  National University of Singapore \\
  danvy@acm.org
}

\maketitle

\runninghead{O. Danvy}{Getting There and Back Again}

\vspace*{-7mm}
\begin{abstract}
``There and Back Again'' (TABA) is a programming pattern where the
recursive calls traverse one data structure and the subsequent returns
traverse another.
This article presents new TABA examples, refines existing ones, and
formalizes both their control flow and their data flow using the Coq
Proof Assistant.
Each formalization mechanizes a pen-and-paper proof, thus making it
easier to ``get'' TABA.
In addition, this article identifies and illustrates a tail-recursive
variant of TABA, There and Forth Again (TAFA)
that does not come back but goes forth instead with more tail calls.

\medskip\noindent
\textbf{Keywords:}
TABA,
recursion pattern,
list processing,
symbolic convolutions,
eta redexes,
Coq Proof Assistant, %
There and Forth Again (TAFA),
defunctionalization and refunctionalization,
lambda-lifting and lambda-dropping,
lightweight fusion and lightweight fission,
continuations.
\end{abstract}

%

\begin{frameit}
\vspace{-4mm}
\begin{quote}
Dear Reader:

Unless you are already acquainted with
``There and Back Again,''
could you first spend a few
minutes thinking about the following programming exercises before
proceeding any further?
Thank you.

\begin{description}[leftmargin=3.5mm]
\vspace*{-2mm}
\item[Convolving a list with itself:] \ \\
Given a list $[x_1, x_2, ..., x_{n-1}, x_n]$, where $n$ is unknown, \\
construct $[(x_1, x_n), (x_2, x_{n-1}),$ $..., (x_{n-1}, x_2), (x_n, x_1)]$ %
in $n$ recursive calls. \\
The implementation should be expressible using a fold function for lists.
\end{description}
\end{quote}
\vspace{0.5mm}
\end{frameit}


\begin{frameit}
\vspace{-4mm}
\begin{quote}
\begin{description}[leftmargin=3.5mm]

\item[Convolving a list with itself (continued):] \ \\
  Here is a non-solution in OCaml:

\inputocamlscnv{SELF_CONVOLVE}

  (This implementation is not a solution because it incurs two independent
  traversals: one for reversing the second list (with
  \inlineml{List.rev}), and one for
  zipping
  the given list and its reverse (with \inlineml{List.map2}).)

\item[Deciding whether two lists are reverses of each other:] \ \\
  Given two lists of unknown lengths,
  test
  whether one list is the reverse of the other, if they have the
  same length $n$.
  Each list should be traversed only once, no
  intermediate list should be
  created, and
  the implementation should proceed in $n$ recursive calls, \ie, be expressible
  using a fold function.
  Here is a non-solution in OCaml:

\inputocamlrevtwo{REV2}

  (This implementation is not a solution because it incurs two independent
  traversals: one for reversing the second list (with \inlineml{List.rev}), and one
  for comparing the first list and the reversed second list
  (with \inlineml{=}).)

\item[Deciding whether a lambda term has the shape of an eta redex:] \ \\
Given the abstract-syntax tree of a $\lambda$ term, test
whether it is shaped like the $\eta$  redex
$\lambda x_1.\lambda x_2.\cdots$ $\lambda x_n.e\:x_1\:x_2\cdots x_n$,
where $e$ and $n$ are unknown,
in $n$ recursive calls.

\item[Indexing a list from right to left:] \ \\
  Given a list and a non-negative integer,
  index this list from the right:
\inputocamlindex{TEST_LIST_INDEX_RTL}

  The list should be traversed only once and no
  intermediate list should be
  created.
  Here are two non-solutions in OCaml:

\inputocamlindex{LIST_NTH_RTL_REV_LIST}
\vspace{-2mm}
\inputocamlindex{LIST_NTH_RTL_REV_INDEX}
\end{description}
\end{quote}
\vspace{-1mm}
\end{frameit}

\begin{frameit}
\vspace{-4mm}
\begin{quote}
\begin{description}[leftmargin=3.5mm]

\item[Indexing a list from right to left (continued):] \ \\
  (The first implementation passes the unit test but is not a solution because it independently
  traverses the given list three times: once for computing its length
  (with \inlineml{List.length}), once for reversing it (with
  \inlineml{List.rev}), and once for indexing it from left to right (with
  \inlineml{List.nth}).
  Also, it constructs an
  intermediate list (with \inlineml{List.rev}).
  The second implementation also passes the unit test but is not a solution either because it
  independently traverses the given list twice: once for computing its
  length and once for indexing it from left to right.)
\end{description}
\end{quote}
\vspace{1.5mm}
\end{frameit}

\section{Background and introduction}
\label{sec:background-and-introduction}

``There and Back Again'' (TABA for short~\cite{Danvy-Goldberg:FI05-shorter}) was a
new programming pattern of structural recursion where one data structure
is traversed at call time (as usual) and another at return time (which
was new). This  pattern makes it possible, for example, to symbolically
convolve the lists
$[v_1; v_2; \ldots; v_{n-1}; v_n]$
and
$[w_1; w_2; \ldots; w_{n-1}; w_n]$
into
$[(v_1, w_n); (v_2, w_{n-1}); \ldots; (v_{n-1}, w_2); (v_n, w_1)]$
in one recursive tra\-versal of either list at call time and one iterative traversal of
the other list at return time, without constructing any intermediate list in reverse order.
Other examples include, \eg, multiplying polynomials, computing Catalan
numbers, testing whether a given list is a palindrome, and testing
whether a given abstract-syntax tree is a generalized
$\beta$ redex without constructing any
intermediate data structure.
All of these examples are therefore expressible using primitive
iteration, \ie, a fold function.

\medskip
The essence of TABA is that traversing a data structure recursively at
call time accumulates enough computing power to traverse another data
structure iteratively at return time.
This second traversal can be either implicit in direct style---which
at first sight may feel as unusual
as seeing Charlie Chaplin or Marcel Marceau walking against the wind
or someone performing the moonwalk dance---or
explicit in
continuation-passing style, where the continuation is delimited and takes
the form of an iterator.
Defunctionalizing this
continuation gives rise to the
intermediate
data structure that is explicitly iterated upon and that one would naturally use in
a tail-recursive solution,
paving  the way to constructing TABA examples
by refunctionalizing non-solutions~\cite{Danvy-Millikin:SCP09-short}.

By now, TABA is often cited as a programming pattern that illustrates
recursion and continuations~\cite{Fernandes-Saraiva:PEPM07,
  Miranda-Perea:ENTCS09, Sergey:PNP14, Shivers-Fisher:JFP06}.
It has inspired a new calculation rule, IO swapping~\cite{Morihata-al:MPC06},
and it has proved useful
to express filters in XML~\cite{Nguyen:PhD},
to illustrate advances in type inference~\cite{Foner:Compose16},
to process tail-aligned lists~\cite{Hemann-Friedman:SFP16},
to illustrate introspection~\cite{Amin-Rompf:POPL18},
and to carry out backpropagation compositionally~\cite{Brunel-al:POPL20,
Wang-al:NIPS18, Wang-al:ICFP19}.
TABA programs have been implemented
in many
languages,
from Prolog and C to
Agda.
At least two have been formalized in Why3~\cite{Filliatre:TABA}.

In this article,
we present new
TABA examples
and we suggest a variation
that simplifies the original treatment of
convolutions
to handle the pathological case where
the two lists to convolve do not have the same length.
This new convolving function traverses both given lists at call time
instead of only one of them and its continuation is only applied when
these two lists have the same length.
The codomain of this continuation
then no longer needs to be an option type to accommodate the pathological case.
The new convolving function is still primitive iterative and thus it can
be expressed using a fold function, whether in direct style or in
continuation-passing style.
We also identify
a tail-recursive variant of TABA, ``There and Forth Again'' (TAFA),
where there is no going back.

All of the above was formalized with the Coq Proof Assistant~\cite{Bertot-Casteran:04}.
For clarity, the proofs use no automation and correspond to what one would
traditionally write by hand~\cite{Bird-Wadler:88, Burstall-Landin:MI69, Manna:74}.
Also, they are all equational, even for implementations that use a continuation.

The rest of this article is structured as follows.
We start by solving the first exercise in the opening ``Dear Reader'' box, namely we
implement a function that convolves any given list of (unknown) length $n$ with
itself in $n$ recursive calls
(\sectionref{sec:scnv}).
We then treat another
simple example of TABA that is also new: given two lists of unknown length, we determine
whether one is the reverse of the other, without reversing either list
(\sectionref{sec:reverse2}).
As for the pathological case where the two lists do not have the same
length, we propose to handle it a priori (\ie, at call time) instead of a
posteriori (\ie, at return time), which simplifies both the program and
its proof (\sectionref{subsec:a-more-perspicuous-solution-of-rev2-where-both-lists-are-first-traversed}
and beyond).
We then revisit the canonical example of TABA,
convolving
two lists
(\sectionref{sec:convolving-lists-when-they-are-supposed-to-have-the-same-length}).
Turning to
the third exercise in the opening ``Dear Reader'' box,
we show how to detect whether a lambda term has the
shape of an eta redex in one recursive descent of its abstract-syntax
tree and without constructing an
intermediate data structure
(\sectionref{sec:detecting-eta-redexes}).
Solving the last exercise in the opening ``Dear Reader'' box suggests a
``There and Forth Again'' programming pattern that is
a tail-recursive variant of TABA in the case where the order in which to
traverse the data structure at return time does not matter
(\sectionref{sec:TAFA}).
We then return to convolving two lists that may not have the same
length
(\sectionref{sec:convolving-lists-when-they-may-not-have-the-same-length})
before concluding (\sectionref{sec:conclusion}) and closing with another
``Dear Reader'' box containing more programming exercises.

\paragraph{Prerequisites and notations:}
This article assumes an elementary knowledge of OCaml and of the Coq
Proof Assistant (and its pure and total functional programming language Gallina).
It also hinges on an awareness of
continuations and defunctionalization, which are reviewed in the appendix.
Notationally, the Haskell plural convention is used throughout: if
\inlineocaml{v} denotes a value, \inlineocaml{vs} denotes a list of
values and if \inlineocaml{p} denotes a pair, \inlineocaml{ps} denotes a
list of pairs.

\section{Convolving a list with itself}
\label{sec:scnv}

The goal of this section is to implement a function that, given a list,
convolves this list with itself.  For example, in OCaml,
this function maps \inlineml{[1; 2; 3; 4]} to \inlineml{[(1, 4); (2, 3);
(3, 2); (4, 1)]}.
The result is a symbolic self-convolution of the given list because
enumerating the first projections of its successive pairs yields the
given list, and enumerating their second projections yields the reverse
of the given list.
(Both enumerations are achieved using the unzip function.
By that book, the zip function achieves a symbolic dot product.)

\medskip
We first program this function in OCaml recursively using the TABA
programming pattern and illustrate this pattern with a trace
(\sectionref{subsec:scnv-first-order-programming}) before formalizing
this implementation using the Coq Proof Assistant
(\sectionref{subsec:scnv-formalizing-and-proving-the-first-order-program}).
We then program this function in OCaml tail recursively with
continuations using the TABA programming pattern and illustrate this
pattern with a trace (\sectionref{subsec:scnv-higher-order-programming})
before formalizing this implementation using the Coq Proof Assistant
(\sectionref{subsec:scnv-formalizing-and-proving-the-higher-order-program}).

\subsection{First-order programming}
\label{subsec:scnv-first-order-programming}

The following implementation uses the TABA programming pattern in that
its auxiliary function traverses the given list (denoted by
\inlineml{vs}) at call time and then traverses the given list (denoted by
\inlineml{ws}) again at return time, constructing the resulting list of
pairs (denoted by \inlineml{ps}):
\inputocamlscnv{SELF_CNV}

\noindent
where ``cnv'' stands for ``convolution''  in the name of this function.

\medskip
Let us visualize the induced computation with a trace:
\begin{itemize}[leftmargin=3.5mm]

\item
  when a function is called, its name and its argument(s) are printed and
  followed by a right arrow;

\item
  when a function returns, its name and its argument(s) are printed and
  followed by a left arrow, itself followed by the result that is being
  returned; and

\item
  the printed names are indented to reflect the nested calls,
  in the tradition of tracing in Lisp.

\end{itemize}

\noindent
To wit:
\inputocamlscnv{A_TRACE_OF_SELF_CNV}

\noindent
In this example, the self-convolution function is called with the list
\inlineml{[1; 2; 3]}, and then calls the visit function recursively as it
traverses this list.
That the given list is traversed is visualized by the successive calls to
the visit function, and that the calls are nested is rendered by the
indentations.
When the end of the list is reached, a pair is returned that contains the
given list and an empty list of pairs, which initiates a series of
returns that matches the series of calls.
That the given list is traversed is visualized by the successive returns
from the visit function, and that the returns match the calls is rendered
by the vertically aligned indentations.
The list of pairs is constructed (1) using the head of the list that was
accessed at call time and that is still available in the lexical
environment and (2) using the head of the returned list.
After the last return, the returned list is empty and the list of pairs
is the symbolic convolution.

\subsection{Formalizing and proving the first-order program}
\label{subsec:scnv-formalizing-and-proving-the-first-order-program}

The implementation above cannot directly be formalized in the Coq Proof
Assistant because of
%
%
the accessors \inlinecoq{List.hd} and \inlinecoq{List.tl} that
optimistically assume that they are applied to a non-empty list
(and otherwise raise an error).
This optimistic assumption actually holds, but the type system is not
strong enough to account for it, and so we have to use an option type,
which incurs a proof obligation that the implementation is total, \ie,
always returns a result constructed with \inlinecoq{Some}.
We also lambda-lift the implementation (see
\appendixref{app:lambda-lifting-and-lambda-dropping}), declaring the
auxiliary function visit globally too instead of
locally, to reason about it using its global name:
\inputcoqscnv{SELF_CNVP}

\noindent
The codomain uses an option type
to account for the impossible case where
the returned list
does not have the same length as the given list (too short above, too
long below):
\inputcoqscnv{SELF_CNV}

\noindent
The list returned by \inlinecoq{self_cnv'},
\inlinecoq{ws},
is supposed to be empty.
The result
is the returned list of pairs, \inlinecoq{ps}.

\medskip
The following theorem captures that the
implementation is sound and complete as well as total:
\inputcoqscnv{SOUNDNESS_AND_COMPLETENESS_OF_SELF_CNV}

\noindent
where \inlinecoq{fst} and \inlinecoq{snd} respectively denote the first
and the second pair projections.
In words -- \inlinecoq{self_cnv} is total in that it always maps a list to
an optional list of pairs that is constructed with \inlinecoq{Some};
it is sound in that enumerating the first projections of the resulting list of
pairs coincides with the given list and enumerating the second
projections of this list of pairs coincides with the reverse of the given
list, which is the definition of a symbolic self-convolution; and it is
complete in that given the symbolic self-convolution \inlinecoq{ps} of a list
\inlinecoq{vs}, applying \inlinecoq{self_cnv} to \inlinecoq{vs} totally
yields \inlinecoq{ps}.

\medskip
This theorem is a corollary of the following lemma about \inlinecoq{self_cnv'}
which captures the essence of TABA's control flow and of TABA's data flow:
\inputcoqscnv{ABOUT_SELF_CNVP}

\noindent
The second argument of \inlinecoq{self_cnv'}, \inlinecoq{vs_sfx}, is the
list traversed by the calls so far and its third argument,
\inlinecoq{vs}, is the given list.
So \inlinecoq{vs_sfx} is a suffix of \inlinecoq{vs}.
In this lemma, this given list is also expressed as the concatenation of a prefix,
\inlinecoq{ws_pfx}, and a suffix, \inlinecoq{ws_sfx}, which are such that
the length of this prefix is the same as the length of the list traversed
so far, \inlinecoq{vs_sfx}.
\begin{description}[leftmargin=3.5mm]

\item[Control flow:]
The lemma expresses how the given list is traversed at return time: the
lengths of \inlinecoq{vs_sfx} and of \inlinecoq{ws_pfx} are the number of
remaining calls to traverse \inlinecoq{vs_sfx}.
By the very nature of structural recursion, this number is also the
number of returns that yield \inlinecoq{Some (ws_sfx, ps)}.
Therefore the returns have traversed the current prefix of the given
list, \inlinecoq{ws_pfx}, and the returned list is its current suffix,
\inlinecoq{ws_sfx}.

\item[Data flow:]
The lemma also captures that the returned list of pairs is
a symbolic convolution of the list that remains to be traversed at call
time, namely \inlinecoq{vs_sfx}, and of the list that has been traversed at
return time, namely \inlinecoq{ws_pfx}.

\end{description}

\noindent
The control-flow aspect of the lemma expresses that \inlinecoq{ws_pfx}
has been traversed by the returns and that \inlinecoq{vs_sfx} remains to
be traversed by the calls.
The data-flow aspect of the lemma expresses that the returned list of
pairs is a symbolic convolution of these two lists.

\medskip
The lemma is proved by induction on \inlinecoq{vs_sfx}.
In the course of this proof, the following property came handy to connect
the first element of a non-empty list and its following suffix and
the last element of this non-empty list and its preceding prefix:
\inputcoqscnv{FROM_FIRST_AND_SUFFIX_TO_PREFIX_AND_LAST}

\noindent
This property is proved by induction on \inlinecoq{vs}.

\subsection{Higher-order programming}
\label{subsec:scnv-higher-order-programming}

The following implementation uses the TABA programming pattern in that
its auxiliary function traverses the given list at tail-call time and
accumulates a two-argument continuation to traverse the given list again and construct
the resulting list of pairs:
\inputocamlscnv{SELF_CNV_C}

Let us visualize the induced computation with a trace:\smallskip
\inputocamlscnv{A_TRACE_OF_SELF_CNV_C}

\noindent
In this example, the self-convolution function is called with the list
\inlineml{[1; 2; 3]}, and then calls the visit function tail-recursively as it
traverses this list and accumulates a continuation.
That the given list is traversed is visualized by the successive calls to
the visit function, and that the calls are tail calls is rendered by the
lack of indentation.
When the end of the list is reached, the current continuation is passed the
given list and an empty list of pairs, and then tail-calls the
accumulated continuations in a way that matches the series of tail-calls
to \inlineml{visit}.
That the given list is traversed is visualized by the successive
tail-calls to the continuations, and that these tail-calls match the
corresponding calls is rendered with the name of these continuations.
The list of pairs is constructed (1) using the head of the list that was
accessed at tail-call time and that is still available in the lexical
environment and (2) using the head of the list that is passed to the
continuation.
When the initial continuation is passed a list and a list of pairs, the former
list is empty and the latter list is the symbolic convolution.

\subsection{Formalizing and proving the higher-order program}
\label{subsec:scnv-formalizing-and-proving-the-higher-order-program}

As in \sectionref{subsec:scnv-formalizing-and-proving-the-first-order-program},
using the partial functions \inlinecoq{List.hd} and \inlinecoq{List.tl}
is not an option in the Coq Proof Assistant, and so we need to use an
option type in the codomain, which also incurs a proof obligation that
the implementation is total. Likewise, the implementation is lambda-lifted and so the local function
is defined globally too, as a recursive equation:
\inputcoqscnv{SELF_CNV_CP}
\inputcoqscnv{SELF_CNV_C}

%
%

\noindent
The initial continuation expects a list and a list of pairs.
This list is supposed to be empty, and the result is this list of pairs.

\medskip
As in \sectionref{subsec:scnv-formalizing-and-proving-the-first-order-program},
the following theorem captures
soundness, completeness, and totality:
\inputcoqscnv{SOUNDNESS_AND_COMPLETENESS_OF_SELF_CNV_C}

Similarly to \sectionref{subsec:scnv-formalizing-and-proving-the-first-order-program},
this theorem is a corollary of the following lemma about \inlinecoq{self_cnv_c'}
which captures the essence of TABA's control flow and of TABA's data flow:
\inputcoqscnv{ABOUT_SELF_CNV_CP}

\noindent
In \sectionref{subsec:scnv-formalizing-and-proving-the-first-order-program},
\inlinecoq{self_cnv'} returns \inlinecoq{Some (ws_sfx, ps)}
whereas \inlinecoq{self_cnv_c'} continues with \inlinecoq{ws_sfx} and
\inlinecoq{ps} here.
This lemma is also proved by induction on \inlinecoq{vs_sfx}.

\subsection{Summary, synthesis, and significance}
\label{subsec:summary-synthesis-and-significance-of-scnv}

Through functional programming and proving, we have illustrated the TABA
programming pattern with one simple example, the symbolic
self-convolution of a list.
The illustration was both visual, using a trace of the successive calls
and returns, and logical, with a lemma that characterizes both the
control flow and the data flow of the TABA programming pattern, be it
recursive or tail-recursive with continuations.

\section{Testing whether two lists are reverses of each other}
\label{sec:reverse2}

The goal of this section is to implement a predicate that decides whether
two given lists of unknown length are reverses of each other.
We are going to inter-derive a
spectrum of
polymorphic functions \inlinecoqfootnotesize{rev2 : forall V : Type, (V -> V -> bool)
  -> list V -> list V -> bool}, each of which satisfies the following
theorem:
\inputcoqrevtwo{SOUNDNESS_AND_COMPLETENESS_OF_REV2}

\noindent
In words -- given a type \inlinecoq{V} of comparable values (and an
associated equality predicate \inlinecoq{beq_V} that is sound and
complete), \inlinecoq{rev2} is sound in that if applying it to two lists
yields \inlinecoq{true}, then the two lists are reverses of each other,
and it is complete in that applying it to two lists that are reverses of
each other yields \inlinecoq{true}.

In each of the implementations, \inlinecoq{rev2} is a main function that
uses one or two auxiliary functions.
Accordingly, the main theorem is a corollary of auxiliary lemmas that we
will state.
Some of these auxiliary lemmas are about the soundness and completeness
of the auxiliary functions, and some others are about a useful property
of these auxiliary functions.

\medskip
We first start from the ``non-solution'' that reverses the first list and
then traverses the reversed first list together with the second list
(\sectionref{subsec:implementation-of-rev2-in-two-passes-that-reverses-the-first-list}).
As it happens, this non-solution is a candidate both for
lightweight fusion (\appendixref{app:lightweight-fusion-and-lightweight-fission})
and for refunctionalization
(\appendixref{app:defunctionalization-and-refunctionalization}):
\vspace{-1mm}
{\let\labelstyle=\textstyle
 \spreaddiagramrows{0.5cm}
 \spreaddiagramcolumns{1.6cm}
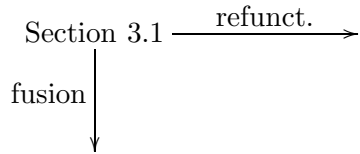
 $$
 \diagram
 {\Text{\sectionref{subsec:implementation-of-rev2-in-two-passes-that-reverses-the-first-list}}}
 \rto<-0.75ex>^{\Text{refunct.}}
 \dto<0ex>_{\Text{fusion}}
 &
 {\phantom{\Text{\sectionref{subsec:implementation-of-rev2-in-two-passes-that-reverses-the-first-list}}}}
 \\
 {\phantom{\Text{\sectionref{subsec:implementation-of-rev2-in-two-passes-that-reverses-the-first-list}}}}
 \enddiagram
 $$
}

\vspace{-4mm}

\noindent
We successively lightweight-fuse it
(\sectionref{subsec:lightweight-fusing-the-implementation-in-two-passes-of-rev2})
and then refunctionalize the result
(\sectionref{subsec:refunctionalizing-the-lightweight-fused-implementation-of-rev2}),
but the converse would do as well, as the end result is the same:
\vspace{-1mm}
{\let\labelstyle=\textstyle
 \spreaddiagramrows{0.5cm}
 \spreaddiagramcolumns{1.6cm}
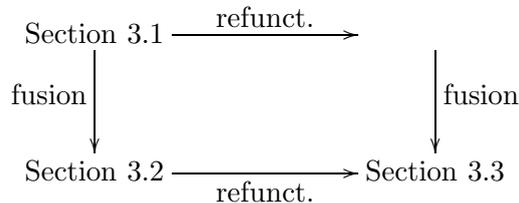
 $$
 \diagram
 {\Text{\sectionref{subsec:implementation-of-rev2-in-two-passes-that-reverses-the-first-list}}}
 \rto<-0.75ex>^{\Text{refunct.}}
 \dto<0ex>_{\Text{fusion}}
 &
 {\phantom{\Text{\sectionref{subsec:implementation-of-rev2-in-two-passes-that-reverses-the-first-list}}}}
 \dto<0ex>^{\Text{fusion}}
 \\
 {\Text{\sectionref{subsec:lightweight-fusing-the-implementation-in-two-passes-of-rev2}}}
 \rto<-0.75ex>_{\Text{refunct.}}
 &
 {\Text{\sectionref{subsec:refunctionalizing-the-lightweight-fused-implementation-of-rev2}}}
 \enddiagram
 $$
}

\vspace{-1mm}

\noindent
As it happens again, this end result is both structurally recursive
and thus expressible using a fold function
(\appendixref{app:generic-programming-with-lists})
and in continuation-passing style (\appendixref{app:continuations-in-a-nutshell}).
We map it back to direct style
(\sectionref{subsec:an-implementation-of-rev2-in-direct-style}) and
then lambda-drop the result
(\appendixref{app:lambda-lifting-and-lambda-dropping}) from two recursive equations to
one block-structured, lexically scoped program
(\sectionref{subsec:an-implementation-of-rev2-in-direct-style-lambda-dropped}).

\medskip
We then turn to the pathological case where the two given lists do not
have the same length,
We simplify this case by treating it at call time rather than at return
time, \ie, by traversing both lists at call time instead of only one of
them, and by only returning if the two given lists have the same length
(\sectionref{subsec:a-more-perspicuous-solution-of-rev2-where-both-lists-are-first-traversed}).
Henceforth, we adjust the block-structured, lexically scoped program, and
then lambda-lift it into two recursive equations, CPS-transform them into
a continuation-passing program, and then defunctionalize and defuse this
program into a non-solution that reverses the first list and then
traverses the reversed first list together with the second list if they
have the same length.
And at each step, we state the corresponding auxiliary lemmas.
Diagrammatically, and with an unambiguous abuse of
notation -- the section numbers are duplicated~\cite{Nolan:Tenet} --
this spectrum can be rendered as follows:

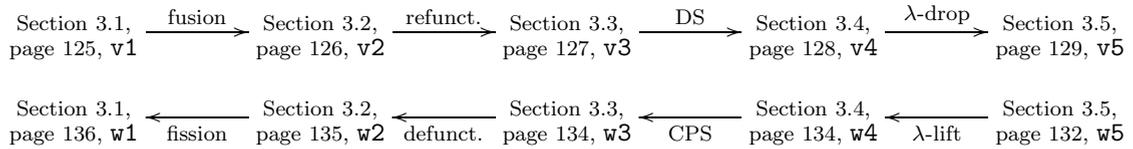
\eject
\hbox{} \vspace{-12mm}

{\scriptsize
{\let\labelstyle=\textstyle
 \spreaddiagramrows{-0.5cm}
 \spreaddiagramcolumns{0.5cm}
 $$
 \diagram
 {\Text{\sectionref{subsec:implementation-of-rev2-in-two-passes-that-reverses-the-first-list},\\
      page~\pageref{subsec:implementation-of-rev2-in-two-passes-that-reverses-the-first-list}, \inlinecoq{v1}}}
 \rto<0ex>^{\Text{fusion}}
 &
 {\Text{\sectionref{subsec:lightweight-fusing-the-implementation-in-two-passes-of-rev2},\\
      page~\pageref{subsec:lightweight-fusing-the-implementation-in-two-passes-of-rev2}, \inlinecoq{v2}}}
 \rto<0ex>^{\Text{refunct.}}%
 &
 {\Text{\sectionref{subsec:refunctionalizing-the-lightweight-fused-implementation-of-rev2},\\
      page~\pageref{subsec:refunctionalizing-the-lightweight-fused-implementation-of-rev2}, \inlinecoq{v3}}}
 \rto<0ex>^{\Text{DS}}
 &
 {\Text{\sectionref{subsec:an-implementation-of-rev2-in-direct-style},\\
      page~\pageref{subsec:an-implementation-of-rev2-in-direct-style}, \inlinecoq{v4}}}
 \rto<0ex>^{\Text{$\lambda$-drop}}
 &
 {\Text{\sectionref{subsec:an-implementation-of-rev2-in-direct-style-lambda-dropped},\\
      page~\pageref{subsec:an-implementation-of-rev2-in-direct-style-lambda-dropped}, \inlinecoq{v5}}}
 \\
 {\Text{\sectionref{subsec:the-new-continuation-passing-implementation-of-rev2-lightweight-fissioned},\\
      page~\pageref{subsec:the-new-continuation-passing-implementation-of-rev2-lightweight-fissioned}, \inlinecoq{w1}}}
 &
 {\Text{\sectionref{subsec:the-new-lightweight-fissioned-implementation-of-rev2-defunctionalized},\\
      page~\pageref{subsec:the-new-lightweight-fissioned-implementation-of-rev2-defunctionalized}, \inlinecoq{w2}}}
 \lto<0ex>^{\Text{fission}}%
 &
 {\Text{\sectionref{subsec:a-new-implementation-of-rev2-in-continuation-passing-style-with-delimited-continuations},\\
      page~\pageref{subsec:a-new-implementation-of-rev2-in-continuation-passing-style-with-delimited-continuations}, \inlinecoq{w3}}}
 \lto<0ex>^{\Text{defunct.}}
 &
 {\Text{\sectionref{subsec:a-new-implementation-of-rev2-in-direct-style-lambda-lifted},\\
      page~\pageref{subsec:a-new-implementation-of-rev2-in-direct-style-lambda-lifted}, \inlinecoq{w4}}}
 \lto<0ex>^{\Text{CPS}}
 &
 {\Text{\sectionref{subsec:a-new-implementation-of-rev2-in-direct-style},\\
      page~\pageref{subsec:a-new-implementation-of-rev2-in-direct-style}, \inlinecoq{w5}}}
 \lto<0ex>^{\Text{$\lambda$-lift}}
 \enddiagram
 $$ }\normalsize
}

In both  Sections~\ref{subsec:lightweight-fusing-the-implementation-in-two-passes-of-rev2},
the program is tail recursive and implements a pushdown automaton where
the intermediate list acts as the stack. In both
Sections~\ref{subsec:refunctionalizing-the-lightweight-fused-implementation-of-rev2},
the program is tail recursive and the stack is represented as an explicit continuation.
And in both
Sections~\ref{subsec:an-implementation-of-rev2-in-direct-style}, the
program is not tail recursive and the explicit continuation is
represented by the implicit control stack that underlies language
processors since Dijkstra~\cite{Dijkstra:60} to implement nested calls in
general and recursive calls in particular, an implementation technique
which is so salient that it is used nowadays to \emph{explain} recursion.
The
continuation-passing programs in
Sections~\ref{subsec:refunctionalizing-the-lightweight-fused-implementation-of-rev2}
and the
direct-style programs in
Sections~\ref{subsec:an-implementation-of-rev2-in-direct-style}
and~\ref{subsec:an-implementation-of-rev2-in-direct-style-lambda-dropped}
also illustrate TABA in that one of the lists is traversed at call
time and the other at return time.
These programs are structurally recursive, or more precisely they are
primitive iterative in that they can be expressed with a fold function:
they can therefore be reasoned about using structural induction.

\subsection{A first-order implementation in two passes}
\label{subsec:implementation-of-rev2-in-two-passes-that-reverses-the-first-list}

Let us start from the non-solution below:
\begin{enumerate}[leftmargin=4.5mm]

\item
  The first list is reversed, tail recursively with an accumulator, using \inlinecoq{rev2'_v1}.

\item
  The reversed first list and the second list are traversed together,
  tail recursively, using \inlinecoq{rev2''_v1}.
  The result is a Boolean: either these two lists have the same length
  and the same elements, \ie, are structurally equal, or they do not.
\end{enumerate}

\noindent
In Gallina:
\inputcoqrevtwo{REVTWOSECOND_VZERO}
\vspace{-1mm}
\inputcoqrevtwo{REVTWOPRIME_VZERO}
\vspace{-1mm}
\inputcoqrevtwo{REVTWO_VZERO}

\eject
\noindent
The two auxiliary definitions give rise to two auxiliary lemmas,
a corollary of which is the soundness and completeness of \inlinecoq{rev2_v1}.
The first is proved by induction on \inlinecoq{vs_op} and the
second by induction on \inlinecoq{vs}:

\inputcoqrevtwo{SOUNDNESS_AND_COMPLETENESS_OF_REVTWOSECOND_VZERO}
\vspace{-1mm}
\inputcoqrevtwo{SOUNDNESS_AND_COMPLETENESS_OF_REVTWOPRIME_VZERO}

\noindent
In words -- \inlinecoq{rev2''_v1} is a sound and complete implementation of
structural equality, and \inlinecoq{rev2'_v1} is a sound and complete
implementation of list reversal with an accumulator.
(To say it again, soundness means that if the implementation yields a
result, this result is correct, and completeness means that if a result
is expected, then the implementation provides it.)

\subsection{A first-order and tail-recursive implementation}
\label{subsec:lightweight-fusing-the-implementation-in-two-passes-of-rev2}

The goal of lightweight fusion by fixed-point promotion~\cite{Ohori-Sasano:POPL07}
is to make an implementation tail recursive, and the implementation in
\sectionref{subsec:implementation-of-rev2-in-two-passes-that-reverses-the-first-list}
is a fitting case for it:
in the definition of \inlinecoq{rev2_v1},
the tail-recursive function \inlinecoq{rev2'_v1} is called,
and its result is used to call \inlinecoq{rev2''_v1}.
Lightweight fusion relocates the context of the call to
\inlinecoq{rev2'_v1} from the definition of \inlinecoq{rev2_v1} to the
definition of \inlinecoq{rev2'_v1}, making it not return,
but instead perform a tail call to \inlinecoq{rev2''_v1}:
\inputcoqrevtwo{REVTWOPRIME_VONE}
\vspace{-1mm}
\inputcoqrevtwo{REVTWO_VONE}

\noindent
In words -- whereas \inlinecoq{rev2'_v1} was called non-tail recursively in
the definition of \inlinecoq{rev2_v1}, which then tail-called
\inlinecoq{rev2''_v1}, \inlinecoq{rev2'_v2} is called tail recursively in
the definition of \inlinecoq{rev2_v2}, which then tail-calls
\inlinecoq{rev2''_v1}.

\medskip
The auxiliary definition above gives rise to an auxiliary lemma,
a corollary of which is the soundness and completeness of \inlinecoq{rev2_v2}:
\inputcoqrevtwo{SOUNDNESS_OF_REVTWOPRIME_VONE}

\noindent
In words -- \inlinecoq{rev2'_v2} ends up tail-calling \inlinecoq{rev2''_v1}
on the reverse of the first list.
This lemma is proved by induction on \inlinecoq{vs}.

\subsection{A higher-order and tail-recursive implementation}
\label{subsec:refunctionalizing-the-lightweight-fused-implementation-of-rev2}

The implementation in
\sectionref{subsec:lightweight-fusing-the-implementation-in-two-passes-of-rev2}
is in defunctionalized form (see \appendixref{app:defunctionalization-and-refunctionalization})
in that the intermediate list (\ie, \inlinecoq{vs_op}), together with
the second pass (\ie, \inlinecoq{rev2''_v1}) that consumes it,
can be represented as a function, \inlinecoq{h_vs_op}:

\inputcoqrevtwo{REVTWOPRIME_VTWO}
\vspace{-1mm}
\inputcoqrevtwo{REVTWO_VTWO}

\noindent
In words -- in
\sectionsref{subsec:implementation-of-rev2-in-two-passes-that-reverses-the-first-list}{subsec:lightweight-fusing-the-implementation-in-two-passes-of-rev2},
the elements of the first list were accumulated into a list in reverse
order (\inlinecoq{vs_op}) and then this list was traversed together with
the second list to check for structural equality in \inlinecoq{rev2''_v1}.
Here, the elements of the first list are accumulated into a function
(\inlinecoq{h_vs_op}) to traverse the second list and then this function
is applied to the second list to carry out this traversal.

\medskip
The auxiliary definition above gives rise to two auxiliary lemmas,
a corollary of which is the soundness and completeness of \inlinecoq{rev2_v3}:
\inputcoqrevtwo{SOUNDNESS_OF_REVTWOPRIME_VTWO}

\noindent
In words -- either \inlinecoq{rev2'_v3} ends up tail-calling
\inlinecoq{h_vs_op} on a suffix of \inlinecoq{ws_given} such that the
corresponding prefix is the reverse of \inlinecoq{vs} or it yields
\inlinecoq{false}.
This soundness lemma is proved by induction on \inlinecoq{vs}.

\inputcoqrevtwo{COMPLETENESS_OF_REVTWOPRIME_VTWO}

\noindent
In words -- if \inlinecoq{vs} is the reverse of a prefix of
\inlinecoq{ws_given}, then \inlinecoq{rev2'_v3} ends up tail-calling
\inlinecoq{h_vs_op} on the corresponding suffix.
This completeness lemma is proved by induction on \inlinecoq{vs}.

Lightweight-fusing Version 1 and then refunctionalizing the result (as
done above) or refunctionalizing Version 1 and then lightweight-fusing
the result yield the same result---and what a remarkable result it is: an
implementation which is
(1) structurally recursive and therefore can be expressed using fold-right
(see \appendixref{app:generic-programming-with-lists}), and
(2) in continuation-passing style (see
\appendixref{app:continuations-in-a-nutshell}) where all calls are tail
calls and \inlinecoq{h_vs_op} acts as the continuation.
This implementation is characteristic of TABA: the first list is
traversed at call time (when the continuation is accumulated) and the
second at return time (when the continuation is applied).

\subsection{A first-order and recursive implementation, lambda-lifted}
\label{subsec:an-implementation-of-rev2-in-direct-style}

In the implementation of
\sectionref{subsec:refunctionalizing-the-lightweight-fused-implementation-of-rev2}
the continuation is delimited: it is initialized in \inlinecoq{rev2_v3},
and it is only applied in \inlinecoq{rev2'_v3}
if the second list
is long enough and its current first element coincides with the corresponding
current first element of the first list; otherwise, the current continuation is
not applied and the computation is discontinued.
In other words, this implementation can be expressed in direct style with
an exception to account for the current continuation not being applied.

\medskip
Gallina being a pure functional language, it does not feature exceptions.
To express this implementation in direct style, we first need to
encode the exceptional behavior using an option type, which makes
the continuation linear (each continuation is defined once and used
once,
last in, first out).
The corresponding direct-style implementation reads as follows:
\inputcoqrevtwo{REVTWOPRIME_VTHREE}
\vspace{-1mm}
\inputcoqrevtwo{REVTWO_VTHREE}

\noindent
This implementation is characteristic of TABA: the successive calls to
\inlinecoq{rev2'_v4} traverse the first list and its successive returns
traverse the second, without creating any intermediate list.
In case of mismatch, \inlinecoq{None} is incrementally returned until the
initial call to \inlinecoq{rev2'_v4} in the definition of
\inlinecoq{rev2_v4}.

\medskip
The auxiliary definition gives rise to two auxiliary lemmas,
a corollary of which is the soundness and completeness of \inlinecoq{rev2_v4}.
These lemmas are proved by induction on \inlinecoq{vs}:
\inputcoqrevtwo{SOUNDNESS_OF_REVTWOPRIME_VTHREE}
\vspace{-1mm}
\inputcoqrevtwo{COMPLETENESS_OF_REVTWOPRIME_VTHREE}

\noindent
In words -- if applying \inlinecoq{rev2'_v4} to \inlinecoq{vs} returns a
list, this list is a suffix of \inlinecoq{ws_given} whose prefix is the
reverse of \inlinecoq{vs}; and if a list is the reverse prefix of
\inlinecoq{ws_given}, applying \inlinecoq{rev2'_v4} to this list
totally yields the corresponding suffix of \inlinecoq{ws_given}.

\medskip
OCaml, however, provides exceptions.
Here is a direct-style implementation in OCaml:
\inputocamlrevtwo{MISMATCHING_LISTS}
\vspace{-1mm}
\inputocamlrevtwo{REVTWOPRIME_VTHREE}
\vspace{-1mm}
\inputocamlrevtwo{REVTWO_VTHREE}

\noindent
The codomain of \inlineocaml{rev2'_v4} is \inlineocaml{bool}, not
\inlineocaml{bool option}, and an exception is raised in case of
mismatch.
Otherwise, the successive calls to \inlinecoq{rev2'_v4} traverse the
first list and its successive returns traverse the second, without
creating any intermediate list, which is the hallmark of TABA.

\subsection{A first-order and recursive implementation, lambda-dropped}
\label{subsec:an-implementation-of-rev2-in-direct-style-lambda-dropped}

As it happens, the implementations
of \sectionref{subsec:an-implementation-of-rev2-in-direct-style} are
recursive equations, \ie, they are in lambda-lifted form (see
\appendixref{app:lambda-lifting-and-lambda-dropping}).
They can be lambda-dropped into a block-structured, lexically scoped program
where the exception is used locally, the auxiliary function is defined locally,
and \inlineocaml{ws_given} occurs free since it is lexically
visible~\cite{Danvy-Schultz:TCS00-short}:
\eject
\inputocamlrevtwo{REVTWO_VFOUR}

\smallskip\noindent
Again, this implementation is structurally recursive and therefore it can
be expressed using fold-right (see
\appendixref{app:generic-programming-with-lists}).
It is also is characteristic of TABA: the first list is traversed at call
time and the second at return time, and no intermediate data structure is
created.

\medskip
Let us visualize the induced computation with traces:
\begin{itemize}[leftmargin=3mm]
\itemsep=0.9pt
\item
In the first example, \inlineocaml{rev2_v5} is called with the lists
\inlineocaml{[1; 2; 3]} and \inlineocaml{[3; 2; 1]}, and then calls
\inlineocaml{rev2'_v5} recursively as it traverses \inlineocaml{[1; 2; 3]}.
When the end of the list is reached, the second list is returned, and the
successive heads of \inlineocaml{[1; 2; 3]} that were accessed at call
time are compared with the successive heads of \inlineocaml{[3; 2; 1]} as
they are accessed at return time, until the initial return from
\inlineocaml{rev2'_v5}.
The final result is \inlineocaml{true} since all the (traced) tests
succeeded and the returned list is empty:
\inputocamlrevtwo{TRACED_REVTWO_VFOUR_1}\vspace{-1mm}

\item
In the second example, \inlineocaml{rev2_v5} is called with the lists
\inlineocaml{[1; 2; 3; 4]} and \inlineocaml{[4; 0; 2; 1]}, and then calls
\inlineocaml{rev2'_v5} recursively as it traverses \inlineocaml{[1; 2; 3; 4]}.
When the end of the list is reached, the second list is returned, and the
successive heads of \inlineocaml{[1; 2; 3; 4]} that were accessed at call
time are compared with the successive heads of \inlineocaml{[4; 0; 2; 1]}
as they are accessed at return time.
The two first heads, \inlineocaml{4}, match, but the two next heads,
\inlineocaml{3} and \inlineocaml{0} do not.
So an exception is raised and the final result is \inlineocaml{false}:

\eject
\hbox{} \vspace*{-10mm}

\inputocamlrevtwo{TRACED_REVTWO_VFOUR_2}\vspace{-1mm}

\item
  The third example illustrates what happens when the second list is
  shorter than the first, namely the remaining returns are skipped:
\inputocamlrevtwo{TRACED_REVTWO_VFOUR_3}\vspace{-1mm}

\item
  The fourth example illustrates what happens when the second list is
  longer than the first, namely the final test fails:
\inputocamlrevtwo{TRACED_REVTWO_VFOUR_4}
\end{itemize}

\vspace*{-4mm}
\subsection{A more perspicuous solution where both lists are first traversed}
\label{subsec:a-more-perspicuous-solution-of-rev2-where-both-lists-are-first-traversed}\vspace{-1mm}

In the implementations above, the job of the first pass (the calls) is to get to the
end of the first list, and the job of the second pass (the returns) is to
traverse the second list, testing whether it is long enough and whether
its successive elements coincide with the corresponding successive elements of the
first list, and then finally testing whether the second list is actually
too long. But there is a simpler algorithm:\vspace{-1mm}
\begin{enumerate}[leftmargin=4.5mm]
\itemsep=0.9pt
\item
  traverse \emph{both} the given lists in the first pass, to establish whether
  they have the same length; and then if they do, and only if they do,
\eject

\item
  traverse the second list, only testing whether its successive elements
  coincide with the corresponding successive elements of the first list,
  knowing that the two lists have the same length.
\end{enumerate}

\noindent
Analysis: the number of recursive calls is the same if the two lists have the same length,
and otherwise this number is the length of the shorter list; also, the second pass
is simpler since it only takes place if the two lists have the same length.

The following sections walk back the path of the previous sections, using
this simpler algorithm and assuming that the reader concurs with its
tenet.

\setcounter{subsection}{4}

\subsection{A more perspicuous first-order and recursive implementation, lambda-dropped}
\label{subsec:a-new-implementation-of-rev2-in-direct-style}

Compared to the implementation in
\sectionref{subsec:an-implementation-of-rev2-in-direct-style-lambda-dropped}
page~\pageref{subsec:an-implementation-of-rev2-in-direct-style-lambda-dropped},
the auxiliary function now traverses both of the given lists and raises
an exception if they do not have the same length:
\inputocamlrevtwo{REVTWO_WFOUR}

\noindent
Since \inlineocaml{rev2'_w5} only returns if the two given lists have the
same length, the nil case in the induction step is impossible, and if the
initial call to \inlineocaml{rev2'_w5} completes, the result is the empty
list.

\medskip
Let us visualize the induced computation with traces,
using the same examples as in
\sectionref{subsec:an-implementation-of-rev2-in-direct-style-lambda-dropped}
page~\pageref{subsec:an-implementation-of-rev2-in-direct-style-lambda-dropped}:

\begin{itemize}[leftmargin=3mm]

\item
In the first example, \inlineocaml{rev2_w5} is called with the lists
\inlineocaml{[1; 2; 3]} and \inlineocaml{[3; 2; 1]}, and then calls
\inlineocaml{rev2'_w5} recursively as it traverses these two lists.
When the end of the two lists is reached, the second list is returned, and the
successive heads of \inlineocaml{[1; 2; 3]} that were accessed at call
time are compared with the successive heads of \inlineocaml{[3; 2; 1]} as
they are accessed at return time, until the initial return from
\inlineocaml{rev2'_w5}: the final result is \inlineocaml{true}:
\inputocamlrevtwo{TRACED_REVTWO_WFOUR_1}

\vspace{-2mm}

\item
In the second example, \inlineocaml{rev2_w5} is called with the lists
\inlineocaml{[1; 2; 3; 4]} and \inlineocaml{[4; 0; 2; 1]}, and then calls
\inlineocaml{rev2'_w5} recursively as it traverses these two lists.
When the end of the two lists is reached, the second list is returned, and the
successive heads of \inlineocaml{[1; 2; 3; 4]} that were accessed at call
time are compared with the successive heads of \inlineocaml{[4; 0; 2; 1]}
as they are accessed at return time.
The two first heads, \inlineocaml{4}, match, but the two next heads,
\inlineocaml{3} and \inlineocaml{0} do not.
So an exception is raised and the final result is \inlineocaml{false}:
\inputocamlrevtwo{TRACED_REVTWO_WFOUR_2}

\vspace{-2mm}

\item
  The third example illustrates what happens when the second list is
  shorter than the first, and is where \inlineocaml{rev2_w5} shines
  compared to \inlineocaml{rev2_v5}.
  The two lists are traversed until the second one reaches the empty
  list.
  Then an exception is raised, and the final result is \inlineocaml{false}:
\inputocamlrevtwo{TRACED_REVTWO_WFOUR_3}

\vspace{-2mm}

\item
  The fourth example illustrates what happens when the second list is
  longer than the first, and is also where \inlineocaml{rev2_w5} shines
  compared to \inlineocaml{rev2_v5}.
  The two lists are traversed until the first one reaches the empty
  list.
  Then an exception is raised, and the final result is \inlineocaml{false}:
\inputocamlrevtwo{TRACED_REVTWO_WFOUR_4}
\end{itemize}

\vspace{-6mm}

\setcounter{subsection}{3}

\subsection{A more perspicuous first-order and recursive implementation, lambda-lifted}
\label{subsec:a-new-implementation-of-rev2-in-direct-style-lambda-lifted}

Lambda-lifting the implementation of
\sectionref{subsec:a-new-implementation-of-rev2-in-direct-style}
page~\pageref{subsec:a-new-implementation-of-rev2-in-direct-style}, yields
two recursive equations:
\inputocamlrevtwo{REVTWOPRIME_WTHREE}
\vspace{-1mm}
\inputocamlrevtwo{REVTWO_WTHREE}

\noindent
The auxiliary function now takes \inlineocaml{ws_given} as an extra parameter
since it is no longer lexically visible in the nil-nil case.

\vspace{-2mm}

\setcounter{subsection}{2}

\subsection{A more perspicuous higher-order and tail-recursive implementation}
\label{subsec:a-new-implementation-of-rev2-in-continuation-passing-style-with-delimited-continuations}

CPS-transforming the implementation of
\sectionref{subsec:a-new-implementation-of-rev2-in-direct-style-lambda-lifted}
page~\pageref{subsec:a-new-implementation-of-rev2-in-direct-style-lambda-lifted},
yields a purely functional program that we can express in Gallina:
\inputcoqrevtwo{REVTWOPRIME_WTWO}
\vspace{-1mm}
\inputcoqrevtwo{REVTWO_WTWO}

\noindent
In words -- \inlinecoq{rev2_w3} accumulates the successive elements of the
first list into a function \inlinecoq{k} to traverse the second list and
compare it to the first one.
Then, if the two lists have the same length, \inlinecoq{k} is
applied to the second list to carry out this comparison.

\medskip
The auxiliary definition above gives rise to two auxiliary lemmas,
a corollary of which is the soundness and completeness of \inlinecoq{rev2_w3}:
\inputcoqrevtwo{SOUNDNESS_OF_REVTWOPRIME_WTWO}

\noindent
In words -- either \inlinecoq{rev2'_w3} ends up tail-calling
\inlinecoq{k} on a suffix of \inlinecoq{ws_given} such that the
corresponding prefix is the reverse of \inlinecoq{vs} or it yields
\inlinecoq{false}.
This soundness lemma is proved by induction on \inlinecoq{vs}.

\inputcoqrevtwo{COMPLETENESS_OF_REVTWOPRIME_WTWO}

\noindent
In words -- if \inlinecoq{vs} is the reverse of a prefix of
\inlinecoq{ws_given}, then \inlinecoq{rev2'_v3} ends up tail-calling
\inlinecoq{k} on the corresponding suffix
if \inlinecoq{vs_given} and \inlinecoq{ws_given} have the same length.
This completeness lemma is proved by induction on \inlinecoq{vs}.

\setcounter{subsection}{1}

\subsection{A more perspicuous first-order and tail-recursive implementation}
\label{subsec:the-new-lightweight-fissioned-implementation-of-rev2-defunctionalized}

The implementation in
\sectionref{subsec:a-new-implementation-of-rev2-in-continuation-passing-style-with-delimited-continuations}
page~\pageref{subsec:a-new-implementation-of-rev2-in-continuation-passing-style-with-delimited-continuations},
is a candidate for defunctionalization, using lists as a data type for
the continuation and \inlinecoq{rev2''_w2} as the corresponding apply
function, which is only called if its arguments have the same length and
which tests their structural equality:
\inputcoqrevtwo{REVTWOSECOND_WONE}
\vspace{-1mm}
\inputcoqrevtwo{REVTWOPRIME_WONE}
\vspace{-1mm}
\inputcoqrevtwo{REVTWO_WONE}

\noindent
Soundness and completeness are proved
by structural induction,
capturing that
the continuation, which has been defunctionalized into \inlinecoq{vs_op} and
\inlinecoq{rev2''_w2}, did implement a comparison between
\inlinecoq{vs_op} and its argument:
\inputcoqrevtwo{SOUNDNESS_AND_COMPLETENESS_OF_REVTWOSECOND_WONE}

\noindent
In words -- \inlinecoq{rev2''_w2} is a sound and complete implementation of
structural equality.

We are now in position to prove a lemma about \inlinecoq{rev2'_w2},
a corollary of which is the soundness and completeness of \inlinecoq{rev2_w2}:
\inputcoqrevtwo{SOUNDNESS_OF_REVTWOPRIME_WONE}

\noindent
In words -- either \inlinecoq{rev2'_w2} ends up tail-calling \inlinecoq{rev2''_w2}
properly and the two given lists have the same length or it does not and
they do not.
This lemma is proved by induction on \inlinecoq{vs}.

\setcounter{subsection}{0}

\subsection{A more perspicuous first-order implementation in two passes}
\label{subsec:the-new-continuation-passing-implementation-of-rev2-lightweight-fissioned}

The implementation in
\sectionref{subsec:the-new-lightweight-fissioned-implementation-of-rev2-defunctionalized}
page~\pageref{subsec:the-new-lightweight-fissioned-implementation-of-rev2-defunctionalized},
is a candidate for lightweight fission by fixed-point demotion (see
\appendixref{app:lightweight-fusion-and-lightweight-fission}) since
\inlinecoq{rev2'_w2} is tail recursive:
\inputcoqrevtwo{REVTWOPRIME_WZERO}
\vspace{-1mm}
\inputcoqrevtwo{REVTWO_WZERO}

\noindent
In words:

\begin{enumerate}[leftmargin=4.5mm]
\itemsep=0.9pt
\item
  Both the first list and the second list are traversed, using
  \inlinecoq{rev2'_w1}, and a reversed copy of the first list is returned
  if the two lists have the same length.

\item
  If the two lists have the same length,
  the reversed first list and the second list are traversed together,
  tail recursively, using \inlinecoq{rev2''_w2}.
  The result is a Boolean: either these two lists have the same elements,
  \ie, are structurally equal, or they do not.
\end{enumerate}

\noindent
We can now prove a lemma about \inlinecoq{rev2'_w1},
a corollary of which is the soundness and completeness of \inlinecoq{rev2_w1}:
\inputcoqrevtwo{SOUNDNESS_OF_REVTWOPRIME_WZERO}

\noindent
In words -- either \inlinecoq{rev2'_w1} yields the reverse of the first given list
and the two given lists have the same length, or it does not and they do not.
This lemma is proved by induction on \inlinecoq{vs}.

\setcounter{subsection}{-1}

\subsection{Summary, synthesis, and significance}
\label{subsec:summary-synthesis-and-significance-of-rev2}

There are many ways to implement a function testing whether two given
lists are reverses of each other.
Each of these ways reflects a particular vision of this computation: tail
recursive with an
intermediate list, tail recursive with a continuation, in
direct style with an exception, or using a fold function.
These implementations form a spectrum in that they can be inter-derived and
thus they all reflect the TABA programming pattern, even the
non-solutions that allocate
intermediate lists.
Being structurally recursive, they can be reasoned about equationally
using structural induction.

\section{Convolving two lists}
\label{sec:convolving-lists-when-they-are-supposed-to-have-the-same-length}

The goal of this section is to implement a function
that symbolically convolves two given lists of unknown length
when these two lists have the same length, which is the original
motivation for TABA~\cite{Danvy-Goldberg:ICFP02}.
As in \sectionref{sec:reverse2}, we are going to inter-derive a spectrum
of implementations of a polymorphic function \inlinecoq{cnv} that
satisfies the following theorem:
\inputcoqcnv{SOUNDNESS_AND_COMPLETENESS_OF_CNV}

\noindent
For brevity, these implementations are written in OCaml below, with a
polymorphic type that reads
\inlineocaml{'a list -> 'b list -> ('a * 'b) list option}.

\medskip
We successively consider implementations in continuation-passing style
(\sectionref{subsec:implementations-in-continuation-passing-style}),
their defunctionalized counterparts
(\sectionref{subsec:first-order-defunctionalized-implementations})
and their lightweight-fissioned counterpart
(\sectionref{subsec:implementations-in-continuation-passing-style-after-lightweight-fission}),
and then the version after both defunctionalization and lightweight fusion in either order
(\sectionref{subsec:first-order-defunctionalized-implementations-after-lightweight-fission}).
We then consider what it takes---\ie, which control operators one needs---to express
the implementations in direct style
(\sectionref{subsec:towards-implementations-in-direct-style}).
Diagrammatically (each of the transformations is reversible):

\vspace{-1mm}
{\let\labelstyle=\textstyle
 \spreaddiagramrows{0.2cm}
 \spreaddiagramcolumns{0.6cm}
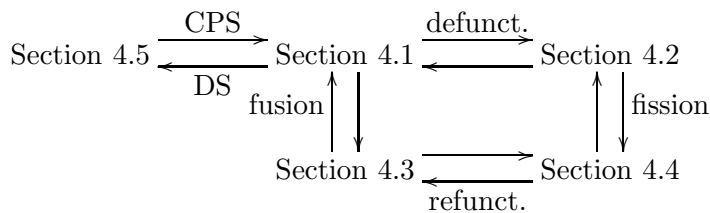
 $$
 \diagram
 {\Text{\sectionref{subsec:towards-implementations-in-direct-style}}}
 \rto<0.5ex>^{\Text{CPS}}
 &
 {\Text{\sectionref{subsec:implementations-in-continuation-passing-style}}}
 \lto<1.5ex>^{\Text{DS}}
 \rto<0.5ex>^{\Text{defunct.}}
 \dto<1ex>
 &
 {\Text{\sectionref{subsec:first-order-defunctionalized-implementations}}}
 \lto<1.5ex>
 \dto<1ex>^{\Text{fission}}
 \\
 &
 {\Text{\sectionref{subsec:implementations-in-continuation-passing-style-after-lightweight-fission}}}
 \rto<0.5ex>
 \uto<1ex>^{\Text{fusion}}
 &
 {\Text{\sectionref{subsec:first-order-defunctionalized-implementations-after-lightweight-fission}}}
 \uto<1ex>
 \lto<1.5ex>^{\Text{refunct.}}
 \enddiagram
 $$
}
\noindent\normalsize
We then turn to the more perspicuous analogue of
\sectionref{subsec:a-more-perspicuous-solution-of-rev2-where-both-lists-are-first-traversed},
\ie, traversing both given lists at call time to determine there and then
whether the two lists have the same length, thus ensuring that there is only a return time
when the two lists have the same length
(\sectionref{subsec:a-more-perspicuous-solution-of-cnv-where-both-lists-are-first-traversed}).
And from then on, \ie, from \sectionref{subsec:more-perspicuous-implementations-in-continuation-passing-style}
to \sectionref{subsec:towards-more-perspicuous-implementations-in-direct-style},
we consider more perspicuous implementations in continuation-passing style,
their refunctionalized or/and lightweight-fissioned counterparts, and what it takes to express
the more perspicuous implementations in direct style, giving rise to a
similar diagram.
\sectionref{subsec:summary-synthesis-and-significance-of-cnv} draws lessons
from these inter-derivations.

\subsection{Implementations in continuation-passing style}
\label{subsec:implementations-in-continuation-passing-style}

This section presents two implementations of the convolution
function using continuations:
\begin{enumerate}[leftmargin=4.5mm]
\itemsep=0.9pt
\item
  one implementation first traverses the first list and
  then returns over the second list;
  during this return, the resulting list of pairs is accumulated tail-recursively
  (the call to \inlinecoq{k} is a tail call);
\item
  the other implementation first traverses the second list and
  then returns over the first list;
  during this return, the resulting list of pairs is constructed recursively
  (the call to \inlinecoq{k} is not a tail call).
\end{enumerate}

\vspace*{-2mm}
\subsubsection{Version that returns over the second list}
\label{subsubsec:implementation-in-continuation-passing-style-that-returns-over-the-second-list}

The first given list is traversed tail-recursively with \inlineocaml{walk},
a continuation
is accumulated,
and eventually this continuation is applied to the second list and to an empty list of pairs,
yielding an optional list of pairs.
The second list is then traversed tail-recursively and a list of pairs is accumulated.
If the two given lists have the same length,
this list of pairs is returned as the result in the initial continuation;
if the second list is shorter than the first,
the computation is discontinued, \ie,
the continuation is not applied and the result is \inlineocaml{None};
and if the second list is longer than the first,
the initial continuation returns \inlineocaml{None}:
\inputocamlcnv{CNV1_CB}

\noindent
This implementation is the motivation for TABA and is due to Goldberg~\cite{Danvy-Goldberg:ICFP02}.
It is structurally recursive and therefore can be expressed using fold-right
and reasoned about using structural induction, since it is also pure
(\ie, uses no computational effects).

\vspace*{-2mm}
\subsubsection{Version that returns over the first list}
\label{subsubsec:implementation-in-continuation-passing-style-that-returns-over-the-first-list}

The second given list is traversed tail-recursively with \inlineocaml{walk},
a continuation
is accumulated,
and eventually this continuation is applied to the first list,
yielding a list of pairs.
The first list is then traversed recursively and a list of pairs is constructed.
If the two given lists have the same length,
this list of pairs is returned as the result;
if the first list is shorter than the second,
the computation is discontinued, \ie,
an exception is raised and the result is \inlineocaml{None};
and if the first list is longer than the second,
an exception is raised in the initial continuation and the result is \inlineocaml{None}:\vspace{1mm}
\inputocamlcnv{CNV2_CB}

\vspace{1mm}\noindent
This implementation is structurally recursive and therefore it can be expressed using fold-right.
It is however impure due to the exception that is used to handle the case where the
first list does not have the same length as the second.
To make it pure, one should change the codomain of the continuation, \eg, using an option type.

 \vspace*{1mm}
\subsection{First-order (defunctionalized) implementations}
\label{subsec:first-order-defunctionalized-implementations}

This section is dedicated to the counterparts of the implementations of
\sectionref{subsec:implementations-in-continuation-passing-style} after
defunctionalization, where the defunctionalized continuation is
represented as an intermediate list
(named either \inlineocaml{xs_op} or \inlineocaml{ys_op} instead of \inlineocaml{k}),
together with a second pass (named \inlineocaml{continue})
that consumes this intermediate list.

\vspace*{1mm}
\subsubsection{Version that returns over the second list}

The first given list is traversed tail-recursively with \inlineocaml{walk},
a list is accumulated in reverse order,
and eventually this reversed list and the second list are traversed in parallel with \inlinecoq{continue},
which traverses the two lists tail-recursively and accumulates a list of pairs.
If the two given lists have the same length, \inlinecoq{continue} returns
this list of pairs as the result:
\inputocamlcnv{CNV1_FO}

\vspace*{-1mm}
\subsubsection{Version that returns over the first list}
\label{subsec:version-that-returns-over-the-first-list-cnv2-fo}

The second given list is traversed tail-recursively with \inlineocaml{walk},
a list is accumulated in reverse order,
and eventually the first list and this reversed list are traversed in parallel with \inlinecoq{continue}.
Both lists are traversed recursively and a list of pairs is constructed.
If the two given lists have the same length, this list of pairs is returned as the result;
otherwise an exception is raised and \inlineocaml{None} is returned.
\inputocamlcnv{CNV2_FO}

\vspace*{-1mm}
\subsection{Implementations in continuation-passing style after lightweight fission}
\label{subsec:implementations-in-continuation-passing-style-after-lightweight-fission}

This section is dedicated to the counterparts of the implementations of
\sectionref{subsec:implementations-in-continuation-passing-style} after
lightweight fission.
Their key point is that the auxiliary function returns the continuation
instead of applying it.

\subsubsection{Version that returns over the second list}

The first given list is traversed tail-recursively with \inlineocaml{walk},
a continuation is accumulated,
and eventually this continuation is returned.
This continuation is then applied to the second list and an empty list of pairs,
the second list is traversed tail-recursively, and a list of pairs is accumulated,
as in \sectionref{subsubsec:implementation-in-continuation-passing-style-that-returns-over-the-second-list}:
\inputocamlcnv{CNV1_CB_LFI}

\noindent
This implementation is structurally recursive and it also fits the pattern of fold-left
(see \appendixref{app:generic-programming-with-lists}):
\inputocamlcnv{CNV1_CB_LFI_LEFT}

\noindent
We note that substituting fold-right for fold-left in this implementation
makes the resulting function implement a dot-product in reverse order,
a consequence of constructing the resulting list of pairs tail-recursively
using an accumulator.
(So did substituting fold-left for fold-right in
\sectionref{subsubsec:implementation-in-continuation-passing-style-that-returns-over-the-second-list}
for the same reason.)

\subsubsection{Version that returns over the first list}
\label{subsec:version-that-returns-over-the-first-list-cnv2-cb-lfi}

The second given list is traversed tail-recursively with \inlineocaml{walk},
a continuation is accumulated,
and eventually this continuation is returned.
This continuation is then applied to the first list,
the first list is traversed recursively, and a list of pairs is constructed,
as in \sectionref{subsubsec:implementation-in-continuation-passing-style-that-returns-over-the-first-list}:
\inputocamlcnv{CNV2_CB_LFI}

\noindent
This implementation is structurally recursive and it also fits the pattern of fold-left:
%
\inputocamlcnv{CNV2_CB_LFI_LEFT}

\noindent
We note that substituting fold-right for fold-left in this implementation
makes the resulting function implement a dot-product,
a consequence of constructing the resulting list of pairs recursively.
(So did substituting fold-left for fold-right in
\sectionref{subsubsec:implementation-in-continuation-passing-style-that-returns-over-the-first-list}
for the same reason.)

\subsection{First-order (defunctionalized) implementations after lightweight fission}
\label{subsec:first-order-defunctionalized-implementations-after-lightweight-fission}

This section is dedicated to the counterparts of the implementations of
\sectionref{subsec:first-order-defunctionalized-implementations} after lightweight fusion,
which are also the counterparts of the implementations of
\sectionref{subsec:implementations-in-continuation-passing-style-after-lightweight-fission}
after defunctionalization.
They construct an
intermediate list as the reverse of one of the two given lists, and then
traverse this reversed list and the other given list.

\subsubsection{Version that returns over the second list}

The first given list is traversed tail-recursively with \inlineocaml{walk},
a list is accumulated in reverse order,
and eventually this list is returned.
This reversed list and the second list are traversed tail-recursively and in parallel with \inlineocaml{continue},
and a list of pairs is accumulated.
If the two given lists have the same length, this list of pairs is returned as the result:\vspace{1mm}
\inputocamlcnv{CNV1_FO_LFI}

\vspace{1mm}
\noindent
The definition of \inlineocaml{walk} coincides with the tail-recursive definition of reverse
that uses an accumulator,
and \inlineocaml{continue} tail-recursively zips together the two lists it is applied to,
using an accumulator.

\subsubsection{Version that returns over the first list}
\label{subsec:version-that-returns-over-the-first-list-cnv2-fo-lfi}

The second given list is traversed tail-recursively with \inlineocaml{walk},
a list is accumulated in reverse order,
and eventually this list is returned.
The first list and this reversed list are traversed recursively and in parallel with \inlinecoq{continue},
and a list of pairs is constructed.
If the two given lists have the same length, this list of pairs is returned as the result:
\inputocamlcnv{CNV2_FO_LFI}

\noindent
The definition of \inlineocaml{walk} coincides with the tail-recursive definition of reverse
that uses an accumulator,
and \inlineocaml{continue} recursively zips together the two lists it is applied to.

\subsection{Towards implementations in direct style}
\label{subsec:towards-implementations-in-direct-style}

This section is dedicated to the direct-style counterpart of the
continuation-passing implementations in
\sectionref{subsec:implementations-in-continuation-passing-style}.
The first one is straightforward (and uses an exception), and the second
is not (it involves delimited-control operators).

\subsubsection{First version: reversing the first list}
\label{subsec:first-version-reversing-the-first-list-cnv1}

Since \inlineocaml{cnv1_cb}, in
\sectionref{subsubsec:implementation-in-continuation-passing-style-that-returns-over-the-second-list},
is tail recursive throughout, it is simple to express it in direct style,
using an exception to handle the case where the current continuation is
not applied:
 \inputocamlcnv{CNV1_DSE}

\subsubsection{Second version: reversing the second list}
\label{subsec:first-version-reversing-the-first-list-cnv2}

In \sectionref{subsubsec:implementation-in-continuation-passing-style-that-returns-over-the-first-list},
the call to the continuation is not a tail call in the induction step.
The definition of \inlineocaml{cnv2_cb} therefore provides yet another case
for the delimited-control operators shift and reset~\cite{Danvy-Filinski:LFP90},
and yet another illustration of the type mismatch between the codomain
of the continuation that has no option, and the domain of answers
that has an option~\cite{Asai:HOSC09, Materzok-Biernacki:ICFP11}.
We choose to let this sleeping dog lie, as awakening it would require
more background material in a way that is unrelated to TABA.

\subsection{A more perspicuous solution where both lists are first traversed}
\label{subsec:a-more-perspicuous-solution-of-cnv-where-both-lists-are-first-traversed}

In the implementations above, the job of the first pass (the calls) is to get to the
end of one of the given lists, and the job of the second pass (the returns) is to
traverse the other given list, checking that it is long enough,
pairing the successive elements of the first list at the point of call
together with the successive elements of the second list at the corresponding point of return,
grouping these pairs into a list,
and then finally testing whether the other list is actually too long.
But as in \sectionref{subsec:a-more-perspicuous-solution-of-rev2-where-both-lists-are-first-traversed}
there is a simpler algorithm:

\begin{enumerate}[leftmargin=4.5mm]
\item
  traverse \emph{both} the given lists in the first pass, to establish whether
  they have the same length; and then if they do, and only if they do,

\item
  traverse the other list to construct the list of pairs,
  knowing that the two given lists have the same length.
\end{enumerate}

\noindent
Analysis: as in \sectionref{subsec:a-more-perspicuous-solution-of-rev2-where-both-lists-are-first-traversed},
the number of recursive calls is the same if the two lists have the same length,
and otherwise this number is the length of the shorter list; also, the second pass
is simpler since it only takes place if the two lists have the same length.

\medskip
The following sections walk through the path of the previous sections,
using this simpler algorithm: 

{\let\labelstyle=\textstyle
 \spreaddiagramrows{0.2cm}
 \spreaddiagramcolumns{0.6cm}
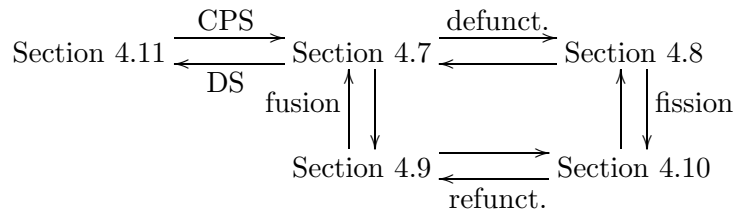
 $$
 \diagram
 {\Text{\sectionref{subsec:towards-more-perspicuous-implementations-in-direct-style}}}
 \rto<0.5ex>^{\Text{CPS}}
 &
 {\Text{\sectionref{subsec:more-perspicuous-implementations-in-continuation-passing-style}}}
 \lto<1.5ex>^{\Text{DS}}
 \rto<0.5ex>^{\Text{defunct.}}
 \dto<1ex>
 &
 {\Text{\sectionref{subsec:more-perspicuous-first-order-defunctionalized-implementations}}}
 \lto<1.5ex>
 \dto<1ex>^{\Text{fission}}
 \\
 &
 {\Text{\sectionref{subsec:more-perspicuous-implementations-in-continuation-passing-style-after-lightweight-fission}}}
 \rto<0.5ex>
 \uto<1ex>^{\Text{fusion}}
 &
 {\Text{\sectionref{subsec:more-perspicuous-first-order-defunctionalized-implementations-after-lightweight-fission}}}
 \uto<1ex>
 \lto<1.5ex>^{\Text{refunct.}}
 \enddiagram
 $$
}

\normalsize \vspace{-3mm}
\subsection{More perspicuous implementations in continuation-passing style}
\label{subsec:more-perspicuous-implementations-in-continuation-passing-style}

This section is dedicated to two implementations of the convolution
function using continuations.
Each of these implementations traverses both lists tail recursively and
in parallel.
If these lists do not have the same length, the result is
\inlineocaml{None}.
Otherwise,

\begin{enumerate}[leftmargin=4.5mm]
\item
  the first implementation returns over the second list;
  during this return, the resulting list of pairs is accumulated tail-recursively
  (the call to \inlinecoq{k} is a tail call);
\item
  the second implementation returns over the first list;
  during this return, the resulting list of pairs is constructed recursively
  (the call to \inlinecoq{k} is not a tail call).
\end{enumerate}

\vspace{-1mm}
\subsubsection{Version that returns over the second list}

Both given lists are traversed tail-recursively
with \inlineocaml{walk},
as a continuation is accumulated.
If the two lists do not have the same length, the continuation is ignored
and the (optional) result is nothing.
Otherwise, the continuation is applied to the second list and to an empty list of pairs.
The second list is then traversed tail-recursively and a list of pairs is accumulated.
The (optional) result is the final version of the accumulator:
\inputocamlcnv{CNW1_CB}

\subsubsection{Version that returns over the first list}

Both given lists are traversed tail-recursively
with \inlineocaml{walk}
as a continuation is accumulated.
If the two lists do not have the same length, the continuation is ignored
and the (optional) result is nothing.
Otherwise, the continuation is applied to the first list.
The first list is then traversed recursively and a list of pairs is constructed.
The (optional) result is this list of pairs:
\inputocamlcnv{CNW2_CB}

\noindent
The nil case in the continuation is impossible, so compared to \inlineocaml{cnv2_cb}
in \sectionref{subsubsec:implementation-in-continuation-passing-style-that-returns-over-the-first-list},
there is no need for an exception.

\subsection{More perspicuous first-order (defunctionalized) implementations}
\label{subsec:more-perspicuous-first-order-defunctionalized-implementations}

This section is dedicated to the counterparts of the implementations of
\sectionref{subsec:more-perspicuous-implementations-in-continuation-passing-style} after
defunctionalization, where the defunctionalized continuation is
represented as an intermediate list
(named either \inlineocaml{xs_op} or \inlineocaml{ys_op} instead of \inlineocaml{k}),
together with the second pass (named \inlineocaml{continue})
that consumes this intermediate list.

\subsubsection{Version that returns over the second list}

Both given lists are traversed tail-recursively
with \inlineocaml{walk}
as the reverse of the first list is accumulated.
If the two lists do not have the same length, the result is \inlineocaml{None}.
Otherwise, the reversed first list and the second list are zipped together tail recursively
with \inlinecoq{continue},
using an accumulator,
and the (optional) result is the final value of this accumulator:
\inputocamlcnv{CNW1_FO}

\subsubsection{Version that returns over the first list}

Both given lists are traversed tail-recursively
with \inlineocaml{walk}
as the reverse of the second list is accumulated.
If the two lists do not have the same length, the result is \inlineocaml{None}.
Otherwise, the first list and the reversed second list are zipped together recursively
with \inlinecoq{continue},
and the (optional) result is the resulting list of pairs:
\inputocamlcnv{CNW2_FO}

\noindent
The nil case in \inlinecoq{continue} is impossible, so compared to \inlineocaml{cnv2_fo}
in \sectionref{subsec:version-that-returns-over-the-first-list-cnv2-fo},
there is no need for an exception.

\subsection{More perspicuous implementations in continuation-passing style after \\ lightweight fission}
\label{subsec:more-perspicuous-implementations-in-continuation-passing-style-after-lightweight-fission}

This section is dedicated to the counterparts of the implementations of
\sectionref{subsec:more-perspicuous-implementations-in-continuation-passing-style} after
lightweight fission.
Their key point is that the auxiliary function returns the continuation
instead of applying it.

\subsubsection{Version that returns over the second list}

Both given lists are traversed tail-recursively
with \inlineocaml{walk},
as a continuation is accumulated.
If the two lists do not have the same length, the continuation is ignored
and the result is \inlineocaml{None}.
Otherwise, the continuation is returned, and is then applied to the second list and an empty list of pairs;
the second list is traversed tail-recursively,
and a list of pairs is accumulated and eventually returned as the result in the initial continuation:\vspace*{1mm}
\inputocamlcnv{CNW1_CB_LFI}

\subsubsection{Version that returns over the first list}

Both given lists are traversed tail-recursively
with \inlineocaml{walk}
as a continuation is accumulated.
If the two lists do not have the same length, the continuation is ignored
and the result is \inlineocaml{None}.
Otherwise, the continuation is applied to the first list, which is traversed recursively,
and a list of pairs is constructed that forms the (optional) result:
\inputocamlcnv{CNW2_CB_LFI}

\noindent
The nil case in the continuation is impossible, so compared to \inlineocaml{cnv2_cb_lfi}
in \sectionref{subsec:version-that-returns-over-the-first-list-cnv2-cb-lfi},
there is no need for an exception.

\subsection{More perspicuous first-order (defunctionalized) implementations after \\ lightweight fission}
\label{subsec:more-perspicuous-first-order-defunctionalized-implementations-after-lightweight-fission}

This section is dedicated to the counterparts of the implementations of
\sectionref{subsec:more-perspicuous-implementations-in-continuation-passing-style-after-lightweight-fission} after
defunctionalization, where the defunctionalized continuation is
represented as an intermediate list
(named either \inlineocaml{xs_op} or \inlineocaml{ys_op} instead of \inlineocaml{k}),
together with the second pass (named \inlineocaml{continue})
that consumes this intermediate list.
Their key point is that they are non-solutions since they construct an
intermediate list as the reverse of one of the two given lists, and then
traverse this reversed list and the other given list.

\subsubsection{Version that returns over the second list}

Both given lists are traversed tail-recursively
with \inlineocaml{walk}
as the reverse of the first list is accumulated.
If the two lists do not have the same length, the result is \inlineocaml{None}.
Otherwise, the reversed first list is returned.
The reversed first list and the second list are then zipped together tail recursively with \inlinecoq{continue},
using an accumulator,
and the (optional) result is the final version of this accumulator: \vspace*{1mm}
\inputocamlcnv{CNW1_FO_LFI}

\vspace*{-3mm}
\subsubsection{Version that returns over the first list}

Both given lists are traversed tail-recursively
with \inlineocaml{walk}
as the reverse of the second list is accumulated.
If the two lists do not have the same length, the result is \inlineocaml{None}.
Otherwise, the reversed second list is returned.
The first list and the reversed second list are then zipped together recursively with \inlinecoq{continue},
and the (optional) result is the resulting list of pairs:  \vspace*{1mm}
\inputocamlcnv{CNW2_FO_LFI}

\vspace*{1mm}\noindent
The nil case in \inlinecoq{continue} is impossible, so compared to \inlineocaml{cnv2_fo_lfi}
in \sectionref{subsec:version-that-returns-over-the-first-list-cnv2-fo-lfi},
there is no need for an exception.

\subsection{Towards more perspicuous implementations in direct style}
\label{subsec:towards-more-perspicuous-implementations-in-direct-style}

This section is dedicated to the direct-style counterpart of the
continuation-passing implementations in
\sectionref{subsec:more-perspicuous-implementations-in-continuation-passing-style}.
The first one is straightforward (and uses an exception), and the second
is not (it involves delimited-control operators).\vspace*{-1mm}

\subsubsection{Version that returns over the second list}

As in \sectionref{subsec:first-version-reversing-the-first-list-cnv1},
it is simple to express \inlineocaml{cnw1_cb} in direct style, using an exception:
\inputocamlcnv{CNW1_DSE}\vspace*{-1mm}

\subsubsection{Version that returns over the first list}

As in \sectionref{subsec:first-version-reversing-the-first-list-cnv2},
the definition of \inlineocaml{cnw2_cb} provides yet another case
for shift and reset, which again we refrain from elaborating.\vspace*{-1mm}

\subsection{Summary, synthesis, and significance}
\label{subsec:summary-synthesis-and-significance-of-cnv}

Throughout, each two versions contrasts the tail-recursive accumulation of the resulting list of pairs
and the recursive construction of this resulting list.
The spectrum of implementations ranges from completely explicit (two passes and an intermediate (reversed) list)
to completely implicit (one recursive descent that is possibly interrupted by raising an exception).
That lightweight fission makes a continuation-passing implementation expressible using fold-left
came as a surprise to the author, and seems new.
The effect of replacing one fold functional by the other crystallizes the relation
between convolving two lists and constructing their dot-product:
a recursive construction yields the resulting list of pairs in the order of the two given lists,
and a tail-recursive accumulation yields the resulting list of pairs in the reverse order.

\medskip
As for reasoning about these implementations (the pure ones, that is),
it involves the same apparatus as in \sectionref{sec:reverse2}:
structural induction and equational reasoning.

%
\section{Deciding whether a lambda term has the shape of an eta redex}
\label{sec:detecting-eta-redexes}

Given a $\lambda$ term, we want to know whether it has the shape
$\elamn
   {x_1}
   {\elamn
      {x_2}
      {\cdots\elamn
               {x_n}
               {\eappn
                  {\eappn
                     {\eappn
                        {e}
                        {x_1}}
                     {x_2}\cdots}
                  {x_n}}}}$,
for some (unknown) expression $e$ and depth $n$.
If it does, we would like to know
this expression and
this depth, in $n$ recursive calls.
We refer to a term of this shape as ``an $\eta$ redex,'' a slight abuse
of terminology since $\lambda x.x\:x$ is not an $\eta$ redex, for
example, but this abuse is inessential here.
We first consider $\lambda$ terms with names
to determine whether they have the shape of
$\elamn
   {x_1}
   {\elamn
      {x_2}
      {\cdots\elamn
               {x_n}
               {\eappn
                  {\eappn
                     {\eappn
                        {e}
                        {x_1}}
                     {x_2}\cdots}
                  {x_n}}}}$
for some $e$ and $n$
(\sectionref{subsec:lambda-terms-with-names}, using OCaml),
then $\lambda$ terms with de Bruijn levels~\cite{de-Bruijn:IM72}
to determine whether they have the shape of
$\elaml
   {\elaml
      {\cdots\elaml
               {\eappl
                  {\eappl
                     {\eappl
                        {e}
                        {0}}
                     {1}\cdots}
                  {(n-1)}}}}$
for some $e$ and $n$
(\sectionref{subsec:lambda-terms-with-de-Bruijn-levels}, using the Coq
Proof Assistant), and then
$\lambda$ terms with de Bruijn indices~\cite{de-Bruijn:IM72}
to determine whether they have the shape of
$\elami
   {\elami
      {\cdots\elami
               {\eappi
                  {\eappi
                     {\eappi
                        {e}
                        {(n - 1)}}
                     {(n - 2)}\cdots}
                  {0}}}}$
for some $e$ and $n$
(\sectionref{subsec:lambda-terms-with-de-Bruijn-indices}, using the Coq
Proof Assistant).

%

\subsection{Lambda terms with names}
\label{subsec:lambda-terms-with-names}

Here is a stylized data type for the abstract-syntax trees of $\lambda$
terms with names, using a unary constructor \inlinecoq{Expn} to stand for
any term that is not a variable, a $\lambda$ abstraction, or an
application:
\inputocamletapn{EXPN}

\noindent
An $\eta$ redex of depth $d$ reads
$\elamn
   {x_1}
   {\elamn
      {x_2}
      {\cdots\elamn
               {x_d}
               {\eappn
                  {\eappn
                     {\eappn
                        {e}
                        {x_1}}
                     {x_2}\cdots}
                  {x_d}}}}$,
but in actuality, this redex contains many implicit parentheses and
actually reads
$\elamn
   {x_1}
   {\elamnp
      {x_2}
      {\cdots\elamnp
               {x_d}
               {\eappnp
                  {(\cdots{\eappnp
                             {\eappnp
                                {e}
                                {x_1}}
                             {x_2}}\cdots)}
                  {x_d}}\cdots}}$.
Reflecting these implicit parentheses, the following function constructs
an $\eta$ redex, given a term and a depth:
\inputocamletapn{MAKE_ETA_REDEXN}

\noindent
In words -- \inlineocaml{visit} recursively constructs the resulting term
with \inlineocaml{Lamn} while accumulating the body around the given
expression with \inlineocaml{Appn}.
Characteristically of $\lambda$ terms with names, a (hopefully) fresh
name is generated for each instance of \inlineocaml{Lamn} using a counter
that is initialized with 1.

\medskip
For example, evaluating \inlineocaml{make_eta_redexn Exp 3} yields
\inputocamletapn{EXAMPLE_OF_ETA_REDEXN_OF_DEPTH_3}

\noindent
Here is its abstract-syntax tree, rotated 90 degrees counterclockwise and flattened:
\vspace{-1mm}
{\let\labelstyle=\textstyle
 \spreaddiagramrows{-0.2cm}
 \spreaddiagramcolumns{-0.2cm}
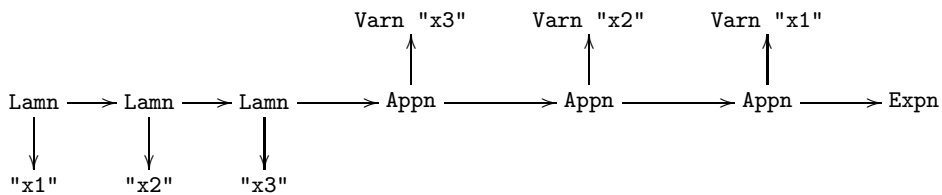
 $$
 \diagram
 &
 &
 &
 {\Text{\footnotesize\texttt{Varn "x3"}}}
 &
 {\Text{\footnotesize\texttt{Varn "x2"}}}
 &
 {\Text{\footnotesize\texttt{Varn "x1"}}}
 &
 \\
 {\Text{\footnotesize\texttt{Lamn}}}
 \rto<-0.6ex>
 \dto
 &
 {\Text{\footnotesize\texttt{Lamn}}}
 \rto<-0.6ex>
 \dto
 &
 {\Text{\footnotesize\texttt{Lamn}}}
 \rto<-0.6ex>
 \dto
 &
 {\Text{\footnotesize\texttt{Appn}}}
 \rto<-0.6ex>
 \uto
 &
 {\Text{\footnotesize\texttt{Appn}}}
 \rto<-0.6ex>
 \uto
 &
 {\Text{\footnotesize\texttt{Appn}}}
 \rto<-0.6ex>
 \uto
 &
 {\Text{\footnotesize\texttt{Expn}}}
 \\
 {\Text{\footnotesize\texttt{"x1"}}}
 &
 {\Text{\footnotesize\texttt{"x2"}}}
 &
 {\Text{\footnotesize\texttt{"x3"}}}
 \enddiagram
 $$
}

Detecting whether a given $\lambda$ term with names has the shape of an $\eta$ redex
fits TABA in that we can recursively descend into its abstract-syntax
tree for as long as we encounter \inlineocaml{Lamn} constructors, until
we encounter another constructor.
At that point, if the tree has the shape of an $\eta$ redex, its depth is
the number of recursive calls that have taken us there.
We can then return the tree and keep traversing it at return time for as
long as we encounter \inlineocaml{Appn} constructors whose second arguments are
a variable whose name matches the name in the corresponding
\inlineocaml{Lamn} constructor.
(This name is visible in the lexical environment of the current recursive call.)
If any other constructor than \inlineocaml{Lamn} (during the calls) and
\inlineocaml{Appn} (during the returns) is encountered, the term is not
an $\eta$ redex.
The term is an $\eta$ redex if the calls and the returns encounter as
many occurrences of \inlineocaml{Lamn} and of \inlineocaml{Appn} and if
the second argument of each occurrence of \inlineocaml{Appn} (\ie, the
actual parameter) is a variable whose name matches the name declared in
the matching \inlineocaml{Lamn} constructor (\ie, the formal parameter).

\medskip
The following implementation is $\lambda$ dropped and uses TABA in direct
style with an option throughout:
\inputocamletapn{ETAPN_DSO}

\noindent
where \inlineocaml{etapn_dso} stands for ``eta predicate for $\lambda$ terms
with names in direct style using an option type.''
At the outset, we check that the depth of the putative $\eta$ redex is
not 0, \ie, is strictly positive.

Let us visualize the induced computation with a trace, rendering the
$\lambda$ terms using the syntax of Scheme:
\inputocamletapn{TRACED_ETAPN_DSO_SOME}

\noindent
The successive \inlineocaml{Lamn} constructors are traversed through the
calls to \inlineocaml{walk} and the successive \inlineocaml{Appn}
constructors are traversed through the subsequent returns, since the
successive names match.

\medskip
Of course most terms are not an $\eta$ redex.
Then the result is \inlineocaml{None}, as illustrated below with Curry's
B combinator:
\inputocamletapn{TRACED_ETAPN_DSO_NONE_B}

To short-circuit the intermediate returns of \inlineocaml{None} if the
given term is not an $\eta$ redex, we can also use an exception:
\inputocamletapn{ETAPN_DSE}
%

\noindent
The codomain of \inlineocaml{walk} is now alleviated to be
\inlineocaml{expn * int} instead of \inlineocaml{(expn * int) option}
since this function only returns if the given term has not proved yet not
to be an $\eta$ redex.
The short circuit can be illustrated with Curry's C combinator:
\inputocamletapn{TRACED_ETAPN_DSE_NONE_C}

\vspace*{-2mm}
\subsection{Lambda terms with de Bruijn levels}
\label{subsec:lambda-terms-with-de-Bruijn-levels}

Here is a stylized data type for the abstract-syntax trees of $\lambda$
terms with de Bruijn levels, using a constant constructor \inlinecoq{Expl} to
stand for any term that is not a variable, a $\lambda$ abstraction, or an
application:
\inputcoqetapl{EXPL}

\noindent
An $\eta$ redex of depth $d$ reads
$\elaml
   {\elaml
      {\cdots\elaml
               {\eappl
                  {\eappl
                     {\eappl
                        {e}
                        {0}}
                     {1}\cdots}
                  {(d-1)}}}}$,
but in actuality, it contains many implicit parentheses and actually reads
$\elaml
   {\elamlp
      {\cdots\elamlp
               {\eapplp
                  {(\cdots{\eapplp
                             {\eapplp
                                {e}
                                {0}}
                             {1}}\cdots)}
                  {(d-1)}}\cdots}}$.
Reflecting these implicit parentheses, the following functions construct
an $\eta$ redex, given a term and a depth:
\inputcoqetapl{MAKE_ETA_REDEXL_AUX}
\vspace{-1mm}
\inputcoqetapl{MAKE_ETA_REDEXL}

\noindent
In words -- \inlineocaml{make_eta_redexl_aux} recursively constructs the
resulting term with \inlineocaml{Laml} while accumulating the body around
the given expression with \inlineocaml{Appl}.
Characteristically of $\lambda$ terms with de Bruijn levels, a counter
accounting for the current level is incremented for each instance of
\inlineocaml{Laml}.
This counter is initialized with 0.

\medskip
For example, evaluating \inlinecoq{make_eta_redexl Expl 3} yields
\inputcoqetapl{EXAMPLE_OF_ETA_REDEXL_OF_DEPTH_3}

\noindent
Here is its abstract-syntax tree, rotated 90 degrees counterclockwise and flattened:

\vspace{-1mm}
{\let\labelstyle=\textstyle
 \spreaddiagramrows{-0.2cm}
 \spreaddiagramcolumns{-0.2cm}
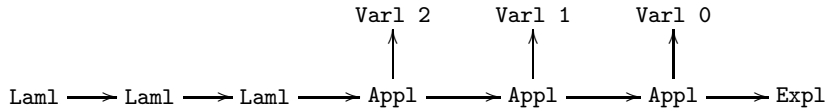
 $$
 \diagram
 &
 &
 &
 {\Text{\footnotesize\texttt{Varl 2}}}
 &
 {\Text{\footnotesize\texttt{Varl 1}}}
 &
 {\Text{\footnotesize\texttt{Varl 0}}}
 &
 \\
 {\Text{\footnotesize\texttt{Laml}}}
 \rto<-0.6ex>
 &
 {\Text{\footnotesize\texttt{Laml}}}
 \rto<-0.6ex>
 &
 {\Text{\footnotesize\texttt{Laml}}}
 \rto<-0.6ex>
 &
 {\Text{\footnotesize\texttt{Appl}}}
 \rto<-0.6ex>
 \uto
 &
 {\Text{\footnotesize\texttt{Appl}}}
 \rto<-0.6ex>
 \uto
 &
 {\Text{\footnotesize\texttt{Appl}}}
 \rto<-0.6ex>
 \uto
 &
 {\Text{\footnotesize\texttt{Expl}}}
 \enddiagram
 $$
}

Detecting whether a given $\lambda$ term with de Bruijn levels has the
shape of an $\eta$ redex fits TABA for the same reason as in
\sectionref{subsec:lambda-terms-with-names}.
So likewise, we traverse the given term recursively for as long as we
encounter \inlinecoq{Laml} constructors, starting with a level 0 and
incrementing this level at each call.
Then we return as soon as we encounter another constructor, and we keep
returning as long as we encounter \inlinecoq{Appl} constructors whose
actual parameter is a variable whose level matches the level in effect
for the matching \inlinecoq{Laml} constructor.
(This level is visible in the lexical environment of the current recursive call.)

The following implementation is $\lambda$ lifted and uses TABA in direct
style with an option throughout, as in
\sectionref{subsec:lambda-terms-with-names}:
\inputcoqetapl{ETAPL_DSP}
\vspace{-1mm}
\inputcoqetapl{ETAPL_DS}

\noindent
where \inlinecoq{etapl_ds} stands for ``eta predicate for $\lambda$ terms
with de Bruijn levels in direct style.''
As befit de Bruijn levels, the level is initialized at the outset and
incremented for each \inlinecoq{Laml} constructor.

\medskip
The soundness and completeness of the implementation are captured by the
following theorem:
\inputcoqetapl{SOUNDNESS_AND_COMPLETENESS_OF_ETAPL_DS}

\noindent
where we wrote \inlinecoq{S d} (\ie, \inlinecoq{d + 1}) since the depth
of an $\eta$ redex is strictly positive.
In words -- \inlinecoq{etapl_ds} is sound in that it totally maps a given
expression \inlinecoq{e} to another expression and a depth, making an
$\eta$ redex with this other expression and that depth yields the given
expression; and it is complete in that given an $\eta$ redex made with a
given expression and a given depth, \inlinecoq{etapl_ds} totally yields
this given expression and this given depth.
This theorem is a corollary of the following lemmas, the first of which
is proved by structural induction over \inlinecoq{e} and the second by
structural induction over \inlinecoq{d}:
\inputcoqetapl{SOUNDNESS_OF_ETAPL_DSP}
\vspace{-1mm}
\inputcoqetapl{COMPLETENESS_OF_ETAPL_DSP}

 \vspace*{-1mm}
\subsection{Lambda terms with de Bruijn indices}
\label{subsec:lambda-terms-with-de-Bruijn-indices}

Here is a stylized data type for the abstract-syntax trees of $\lambda$
terms, using de Bruijn indices (\ie, lexical offsets) and a constant
constructor \inlinecoq{Expi} to stand for any term that is not a
variable, a $\lambda$ abstraction, or an application:
\inputcoqetapi{EXPI}

\noindent
An $\eta$ redex of depth $d$ reads
$\elami
   {\elami
      {\cdots\elami
               {\eappi
                  {\eappi
                     {\eappi
                        {e}
                        {(d - 1)}}
                     {(d - 2)}\cdots}
                  {0}}}}$,
but in actuality, it contains many implicit parentheses and actually reads
$\elami
   {\elamip
      {\cdots\elamip
               {\eappip
                  {(\cdots{\eappip
                             {\eappip
                                {e}
                                {(d - 1)}}
                             {(d - 2)}}\cdots)}
                  {0}}\cdots}}$.
Reflecting these implicit parentheses, the following function constructs
an $\eta$ redex, given a term and a depth:
\inputcoqetapi{MAKE_ETA_REDEXI}

\noindent
In words -- \inlineocaml{make_eta_redexi} recursively constructs the
resulting term with \inlineocaml{Lami} while accumulating the body around
the given expression with \inlineocaml{Appi}.
Characteristically of $\lambda$ terms with de Bruijn indices, the
decreasing counter that was initialized with the desired depth can serve
as the index for the successive indices of each variable in the resulting
$\eta$ redex.

\smallskip
For example, evaluating \inlinecoq{make_eta_redexi Expi 3} yields
\inputcoqetapi{EXAMPLE_OF_ETA_REDEXI_OF_DEPTH_3}
Here is its abstract-syntax tree, rotated 90 degrees counterclockwise and flattened:

\vspace{-1mm}
{\let\labelstyle=\textstyle
 \spreaddiagramrows{-0.2cm}
 \spreaddiagramcolumns{-0.2cm}
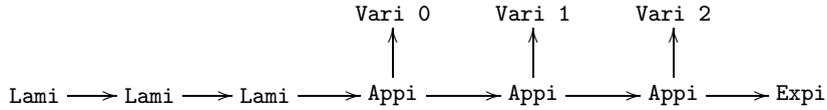
 $$
 \diagram
 &
 &
 &
 {\Text{\footnotesize\texttt{Vari 0}}}
 &
 {\Text{\footnotesize\texttt{Vari 1}}}
 &
 {\Text{\footnotesize\texttt{Vari 2}}}
 &
 \\
 {\Text{\footnotesize\texttt{Lami}}}
 \rto<-0.6ex>
 &
 {\Text{\footnotesize\texttt{Lami}}}
 \rto<-0.6ex>
 &
 {\Text{\footnotesize\texttt{Lami}}}
 \rto<-0.6ex>
 &
 {\Text{\footnotesize\texttt{Appi}}}
 \rto<-0.6ex>
 \uto
 &
 {\Text{\footnotesize\texttt{Appi}}}
 \rto<-0.6ex>
 \uto
 &
 {\Text{\footnotesize\texttt{Appi}}}
 \rto<-0.6ex>
 \uto
 &
 {\Text{\footnotesize\texttt{Expi}}}
 \enddiagram
 $$
}

\vspace*{-1mm}
Detecting whether a given $\lambda$ term with de Bruijn indices has the
shape of an $\eta$ redex fits TABA for the same reason as in
\sectionsref{subsec:lambda-terms-with-names}{subsec:lambda-terms-with-de-Bruijn-levels}.
So likewise, we traverse the given term recursively for as long as we
encounter \inlinecoq{Laml} constructors, and then return the term that
is not constructed with \inlinecoq{Laml} and a counter initialized with 0.
Then we keep returning as long as we encounter \inlinecoq{Appl}
constructors whose actual parameter is a variable whose index matches the
increasing counter.
Eventually, if all tests have succeeded, the result is the expression
around which the $\eta$ redex was constructed together with the depth of
this $\eta$ redex, which must be strictly positive.

\medskip
The following implementation is $\lambda$ lifted so that we can refer to
the auxiliary function in lemmas.
It uses TABA and a
continuation to stop immediately if the given term is not an $\eta$ redex: 
\inputcoqetapi{ETAPI_CBPRIME}
\vspace{-1mm}
\inputcoqetapi{ETAPI_CB}
%

\noindent
where \inlinecoq{etapi_cb} stands for ``continuation-based eta predicate
for $\lambda$ terms with de Bruijn indices.''
In words -- given an $\eta$ redex of depth $n$, \inlinecoq{etapi_cb'} calls
itself recursively $n$ times on the successive \inlinecoq{Lami}
constructors, accumulating a continuation to traverse nested applications
as long as their argument is a variable whose de Bruijn index increases
with the nesting of applications.
The initial continuation is eventually applied to the inner expression
in position of a function
and to the depth of its nesting.

\smallskip
The soundness and completeness of the implementation are captured by the
following theorem:
\inputcoqetapi{SOUNDNESS_AND_COMPLETENESS_OF_ETAPI_CB}

\noindent
where again we wrote \inlinecoq{S d}
since the depth of an $\eta$ redex is strictly positive.
This theorem is is a corollary of the following lemmas, the first of which
is proved by structural induction over \inlinecoq{e} and the second by
structural induction over \inlinecoq{d}:
\inputcoqetapi{SOUNDNESS_OF_ETAPI_CBPRIME}
\vspace{-1mm}
\inputcoqetapi{COMPLETENESS_OF_ETAPI_CBPRIME}

An interesting aspect of this implementation is that its continuations
have no free variables.
Therefore defunctionalizing them yields a data type that is
isomorphic to Peano numbers, and an apply function that iterates over
a given Peano number.
All in all, the result is a
remarkably
simple first-order
tail-recursive predicate
that foreshadows the tail-recursive variant of TABA
introduced in the next section:
\inputcoqetapi{ETAPISECOND}
\vspace{-1mm}
\inputcoqetapi{ETAPIPRIME}
\vspace{-1mm}
\inputcoqetapi{ETAPI}

\noindent
In words -- given an expression and a counter initialized with 0,
\inlinecoq{etapi'} traverses this expression tail-recursively for as long
as it encounters \inlinecoq{Lami} constructors, incrementing the counter
as it goes; then it tail-calls \inlinecoq{etapi''} with the counter, the
sub-expression that does not start with the \inlinecoq{Lami} constructor,
and an index initialized with 0;
given this counter, this sub-expression, and this index,
\inlinecoq{etapi''} tail-recursively traverses the sub-expression for as
long as it encounters \inlinecoq{Appi} constructors whose actual
parameter is a variable whose de Bruijn index matches the current index,
until the counter reaches 0; at that point, \inlinecoq{etapi''} checks
that the resulting depth is strictly positive.

\medskip
The soundness and completeness of this implementation are stated and
proved mutatis mutandis: \vspace{1mm}
\inputcoqetapi{SOUNDNESS_AND_COMPLETENESS_OF_ETAPI}

\vspace{1mm}\noindent
The cognoscenti will have identified that this first-order predicate implements
a counter automaton.
In the same vein, CPS-transforming the predicate for $\lambda$
terms with names from \sectionref{subsec:lambda-terms-with-names} (see
\appendixref{subapp:direct-style-vs-continuation-passing-style}),
splitting its continuation into two (see
\appendixref{subapp:splitting-continuations}), and dropping the one that
corresponds to the \inlinecoq{None} case (see
\appendicesref{app:lambda-lifting-and-lambda-dropping}{subapp:splitting-continuations})
yield the implementation of a pushdown automaton with a stack of names.
Small world, many disguises.

\section{There and forth again (TAFA)}
\label{sec:TAFA}

In some cases, once one gets there, there is no need to go back again:
one can go forth instead to complete the computation, which makes it tail
recursive throughout.
This section illustrates two examples of this situation: the first is the
closing exercise in the ``Dear Reader'' box before \sectionref{sec:background-and-introduction},
and the second is due to Hemann and Friedman~\cite{Hemann-Friedman:SFP16}.

\subsection{Indexing a list from the right}
\label{subsec:indexing-a-list-from-the-right}

Indexing a list from the right means that given an index $n$ and a
list constructed as the concatenation of any list, a singleton list
containing $v$, and any list of length $n$, the result should be $v$.
If the given list is too short for the given index, the result is
undefined.

\subsubsection{Programming}
\label{subsubsec:list-index-rtl-programming}

At first glance, indexing a list from the right provides another
illustration of TABA: traverse the list at call time and eventually
return an intermediate result initialized with the index, using the
\inlineocaml{Index} constructor below; and then at return time, decrement
this index:

\begin{itemize}[leftmargin=3.5mm]
\item
  if the decremented index reaches 0 before the last return, the list is
  long enough and the intermediate result becomes the head of the
  current list, using the \inlineocaml{Found} constructor below;

\item
  if the decremented index has not reached 0 at the last return, the
  list is too short for the index.
\end{itemize}

 \eject
\noindent
To wit:
\inputocamlindex{INTERMEDIATE_RESULT}
\vspace{-1mm}
\inputocamlindex{LIST_INDEX_RTL_DS}

\noindent
To illustrate, here are two traces of the computation, one where the list
is too short for the index, and then one where it is long enough:
\inputocamlindex{TRACED_LIST_INDEX_RTL_DS}

\noindent
In words -- the given list is traversed all the way and at return time, the
index is decremented.
If the index reaches 0 in the course of the returns, the value to index
exists and is returned.

\medskip
On second thought, though, one could use the idea of
\sectionref{subsec:a-more-perspicuous-solution-of-rev2-where-both-lists-are-first-traversed}
and decrement the given index as the given list is traversed:
\begin{itemize}[leftmargin=3.5mm]

\item
  if the end of the list is reached before the index reaches 0, the list
  is too short for the index and the computation can stop;

\item
  conversely, if 0 is reached first, then the list is long enough, and
  the length of the prefix of the given list down to the current suffix
  coincides with the given index;
  all one needs to do then is to go forth and slide through both the
  given list and the current suffix of the given list, preserving this
  length property;
  when the current suffix becomes its end, \ie, the empty list, the
  length property still holds and so the result is the head of
  the current list.
\end{itemize}

\noindent
To wit:
\inputocamlindex{LIST_INDEX_RTL_TAFA}

\smallskip\noindent
As one can see, given a list of length $n$, this implementation operates
in $n$ tail-recursive calls.
(The initial tail call is not recursive.)
To illustrate, here is a trace of the computation:\smallskip
\inputocamlindex{TRACED_LIST_INDEX_RTL_TAFA}

\vspace{-3mm}
\subsubsection{Formalizing and proving}
\label{subsubsec:list-index-rtl-formalizing-and-proving}

Let us formalize this instance of TAFA.
In a nutshell,
\begin{itemize}[leftmargin=3.5mm]
\item
  if the list is empty, it contains no element to index: the list is
  too short, no matter the index;
\item
  if the given list is non-empty, its tail should contain at most as many
  elements as the given index; otherwise, the list is still too short;
\item
  this tail can be traversed as the index is decremented until
  it reaches 0;
  once we get there, the difference between the tail of the given list
  and the current suffix has the same length as the given index.
\end{itemize}

\noindent
We are therefore better off to first test whether the list is empty and
then proceed by induction on the tail of the list if it is non-empty, a
byproduct of thinking before proving, if not before programming.

 \eject
Concretely:
\inputcoqindex{LIST_INDEX_RTL_TAFA_THERE}
\vspace{-1mm}
\inputcoqindex{SOUNDNESS_AND_COMPLETENESS_OF_LIST_INDEX_RTL_TAFA_THERE}

\noindent
In words -- if the given list is not empty,
\inlineocaml{list_index_rtl_there}
traverses its tail,
\inlinecoq{vs_sfx}, and the given index, \inlinecoq{n}, and returns the
\inlinecoq{n}th suffix of this tail, if one exists.
(The \inlinecoq{0}th suffix of a list is itself.)
The lemma is proved by structural induction over the given index.

\medskip
So, after calling \inlinecoq{list_index_rtl_there}, we have access
to the first element of the given list, the tail of the given list, and
the \inlinecoq{n}th suffix of the tail of the given list, where
\inlinecoq{n} is the given index.
We can then go forth and traverse the tail of the given list and the
\inlinecoq{n}th suffix until this suffix becomes the empty list.
At that point, the tail has length \inlinecoq{n} and the element just
before is the element we were looking for.

 \medskip
Concretely:
\inputcoqindex{LIST_INDEX_RTL_TAFA_FORTH}

\noindent
The codomain is an option type because the type system does not guarantee
that traversing the tail of the given list never reaches the empty list.
However, this function is total:
\inputcoqindex{SOUNDNESS_AND_COMPLETENESS_OF_LIST_INDEX_RTL_TAFA_FORTH}

 \eject

\noindent
In words -- given a non-empty list \inlinecoq{v :: vs'} and an index
\inlinecoq{n} that is smaller than the length of \inlinecoq{vs'}, let
\inlinecoq{vs_pfx} denote the prefix of \inlinecoq{vs'} of length
\inlinecoq{n}, and \inlinecoq{vs_sfx} denote the corresponding suffix.
Since \inlinecoq{n} is smaller than the length of \inlinecoq{vs'}, an
element \inlinecoq{w} exists in \inlinecoq{vs'} at index \inlinecoq{n}
going from right to left and so does the suffix \inlinecoq{ws_sfx} that
follows it in \inlinecoq{vs'}, a suffix that has length \inlinecoq{n}.
This element \inlinecoq{w} and this suffix \inlinecoq{ws_sfx} are
computed by \inlinecoq{list_index_rtl_forth} when it is applied to
\inlinecoq{v}, \inlinecoq{vs'}, and \inlinecoq{vs_sfx}.
The lemma is proved by structural induction over \inlinecoq{vs_sfx}.

 \medskip
All told, the opening description of the present section is formalized as follows:
\inputcoqindex{LIST_INDEX_RTL_TAFA}

\noindent
In words -- given the empty list and an index,
\inlinecoq{list_index_rtl_tafa} returns \inlinecoq{None}; given a
non-empty list and an index, it determines whether the \inlinecoq{n}th
suffix of the tail of the given list exists; if it does not,
\inlinecoq{list_index_rtl_tafa} returns \inlinecoq{None}; if it does,
\inlinecoq{list_index_rtl_tafa} computes the suffix of the given list
that has length \inlinecoq{n} and the element that precedes this suffix
(both exist); this element is then the result of indexing the given list
at the given index.
The ``trailing'' pointer always points to a non-empty list (a type property~\cite{Foner:Compose16}).

 \medskip
Soundness and completeness are a corollary of the previous lemmas:
\inputcoqindex{SOUNDNESS_AND_COMPLETENESS_OF_LIST_INDEX_RTL_TAFA}

\subsection{Computing the common suffix of two lists}
\label{subsec:computing-the-common-suffix-of-two-lists}

\lstset{
  rangeprefix=\;\{,
  rangesuffix=\}
}

At the 2016 Scheme Workshop~\cite{Hemann-Friedman:SFP16}, Hemann and
Friedman used TABA to compute the common suffix of two lists of unknown
length in a number of calls and returns that is at most the sum of the
lengths of these two lists.
This computation also fits the TAFA (There and Forth Again) pattern in
that the two lists can be traversed in sync (there), the longest can be slided
through to reach a suffix that has the same length as the shortest (forth), and
the resulting suffix and the shortest list can then be traversed in sync
for comparison (again), using a total number of tail-recursive calls that is
precisely the sum of the lengths of the two given lists.
To ease the comparison with Hemann and Friedman's solution, the programs
in this section are expressed in Chez Scheme~\cite{Dybvig:ICFP06-short}.

\subsubsection{Finding the common suffix of two proper lists that have the same length}
\label{subsubsec:finding-the-common-suffix-of-two-proper-lists-that-have-the-same-length}

Given two proper lists (\ie, lists ending with nil) that have the same
length, one can traverse them in parallel with an outer loop
(\inlinelisp{common-suffix_same-length}) and with an inner loop
(\inlinelisp{inner-loop}).
In the outer loop, \inlinelisp{vs} and \inlinelisp{ws} denote
the current suffix candidate, and in the inner loop, \inlinelisp{vs_sfx} and
\inlinelisp{ws_sfx} denote their respective suffixes.
(Symmetrically, we say that \inlinelisp{vs} denotes a ``trail'' of
\inlinelisp{vs_sfx} and that \inlinelisp{ws} denotes a trail of
\inlinelisp{ws_sfx}.)
If this suffix is empty,
then their trails
both denote the resulting common suffix.
Otherwise, if the heads of the respective suffixes are equal, the inner
loop continues, and if they are not, a new iteration of the outer loop is
initiated with two new trails, \ie, suffix candidates:
\inputlongestcommonsuffix{COMMON-SUFFIX_SAME-LENGTH}

\noindent
For example, here is a trace of this traversal:
\inputlongestcommonsuffix{TRACED-COMMON-SUFFIX_SAME-LENGTH}

\noindent
In words -- the outer loop is initiated with trails \inlinelisp{(1 2 1 2
  3)} and \inlinelisp{(1 2 3 1 2 3)}.
The inner loop traverses their suffix until it encounters two differing
heads (namely \inlinelisp{1} and \inlinelisp{3}).
The outer loop is then re-initiated with trails \inlinelisp{(2 3)} and
\inlinelisp{(2 3)}.
The inner loop traverses their suffix to their end.
The result is either of the current trails.

\subsubsection{Traversing the suffix of a proper list to find a suffix of this list}
\label{subsubsec:traversing-the-suffix-of-a-proper-list-to-find-a-suffix-of-this-list}

The following procedure is given a proper list (of length $m$) and one of
its suffixes (of length $n$), \ie, a suffix and a trail of this suffix.
The procedure traverses them in parallel until the end of the suffix.
It then returns the trail, which is a suffix of the given list that has
length $m - n$:
\inputlongestcommonsuffix{SLIDE}

\noindent
For example, here is a trace of this traversal:
\inputlongestcommonsuffix{TRACED-SLIDE}

\noindent
In words -- \inlinelisp{slide} is applied to a list of length $5$ and to a
suffix of this list that has length $2$; it traverses them both in parallel
until the end of the second list; the first list then has length $5 - 2 =
3$ and is returned.

\subsubsection{Finding a suffix of the longer list with the same length as the shorter list}
\label{subsubsec:finding-a-suffix-of-the-longer-list-with-the-same-length-as-the-shorter-list}

Given two proper lists, we can enumerate their successive suffixes by
traversing them in parallel:

\begin{itemize}[leftmargin=3.5mm]

\item
If both traversals end with nil, the two given lists have the same
length.
We can find their common suffix using \inlinelisp{common-suffix_same-length}.

\item
If one traversal ends with nil but not the other, the two lists do not
have the same length.
We can use \inlinelisp{slide} on the longest list and its current suffix
to obtain a suffix that has the same length as the shortest list,
and then \inlinelisp{common-suffix_same-length} to find their common
suffix.

\end{itemize}

Concretely:
\inputlongestcommonsuffix{LONGEST-COMMON-SUFFIX}

\noindent
For example, here is a trace of this traversal:
\inputlongestcommonsuffix{TRACED-LONGEST-COMMON-SUFFIX}

\noindent
In words -- the two lists (here: \inlinelisp{(2 3 4 5 6 8)} and
\inlinelisp{(3 0 5 0 8)})
are first traversed in parallel to determine
that the first one is longer; what remains of the first list is a
non-empty suffix of it (here: \inlinelisp{(8)}); the first list and its suffix are then slided
across to compute the suffix of the first list that has the same length
as the second list (here: \inlinelisp{(3 4 5 6 8)}); this suffix and the second list are then traversed in
parallel to compute their longest common suffix.
All told, and keeping in mind that initial calls are not recursive,
the number of tail-recursive calls is 11, which is the sum of the
lengths of the two given lists.
(The author has automated this measure and verified it in practice with a
variety of tests, a large number of which involved randomly generated lists.)

\subsubsection{All in all}
\label{subsubsec:all-in-all}

An iterative solution exists for finding the common suffix of two proper
lists of arbitrary length where the number of tail-recursive calls
(initial calls do not count, only recursive ones)
is exactly the sum of the lengths of the two given lists:

\begin{description}[leftmargin=3.5mm]

\item[There:]
  The idea is to traverse both lists in parallel until one of them is nil, which gets us there.
  If both are nil, then the two given lists have the same length; go to Again with both given lists.
  Otherwise, go to Forth with the longer list and its non-empty suffix,
  \ie, with the non-empty suffix and its trail.

\item[Forth:]
  The idea is to traverse both the given suffix list and its trail in
  parallel until the end of this suffix.
  The trail is then a suffix of the given longer list that has the same
  length as the given shorter list; go to Again with this trail and with
  this shorter list.

\item[Again:]
  At that point, both lists have the same length and we can traverse them in parallel with two trail pointers,
  resuming the Again step with new trails if the current heads of the lists are not the same.
  When we reach nil, each of the trails is the common suffix.

\end{description}

\noindent
This instance of ``There and Forth Again'' is an optimization of ``There and Back Again''
that is not always applicable.
When it does, though, it yields a tail-recursive solution.
Here, this optimization is applicable because the order of comparisons in
the list (whether from the end to the beginning or from the beginning to
the end) does not matter.

\lstset{
  rangeprefix=\(*\ \{,
  rangesuffix=\}\ *\)
}

\section{Convolving lists that may not have the same length}
\label{sec:convolving-lists-when-they-may-not-have-the-same-length}

Sometimes, it is not a mistake to convolve lists that do not have
the same length, \eg, to multiply
polynomials~\cite[Section~3]{Danvy-Goldberg:FI05-shorter}, to compute
Catalan numbers~\cite[Section~5]{Danvy-Goldberg:FI05-shorter},
or to compute the Cartesian product of two sets in breadth-first order
rather than the usual depth-first order~\cite{Barron-Strachey:PNNC66}:
one may want to convolve the shorter list with a prefix of the longer
list or with a suffix of it.
Since the first list determines the control flow of the convolving
function, does
one needs to revert to the iterative solution
that constructs an
intermediate list?  No.

\subsection{Convolving a list and the prefix of a longer list (and vice versa)}
\label{subsec:convolving-prefixes}

Let \inlinecoq{vs} and \inlinecoq{ws} denote two lists that have the same
length, and let \inlinecoq{xs} denote another list.
\begin{itemize}[leftmargin=3.5mm]

\item
  Should convolving \inlinecoq{vs} and \inlinecoq{ws ++ xs} reduce to
  convolving \inlinecoq{vs} and \inlinecoq{ws}, then the initial
  continuation ends up being applied to \inlinecoq{xs} and a list of pairs,
  and it should ignore the former and return the latter.

\item
  Should convolving \inlinecoq{vs ++ xs} and \inlinecoq{ws} reduce to
  convolving \inlinecoq{vs} and \inlinecoq{ws}, then the traversal of
  both lists should stop when reaching the end of the second list.
\end{itemize}

\noindent
And indeed,
convolving \inlineml{[1; 2; 3; 4]} and \inlineml{[10; 20]}
and
convolving \inlineml{[1; 2]} and \inlineml{[10; 20; 30, 40]}
yield the same result as
convolving \inlineml{[1; 2]} and \inlineml{[10; 20]} in this case:
the suffix of the longer list is ignored, the convolving function is structurally
recursive, and true to TABA
the convolving function traverses the first list at call time
and the second at return time.

\subsection{Convolving a list and the suffix of a longer list (and vice versa)}
\label{subsec:convolving-suffixes}

Let \inlinecoq{vs} and \inlinecoq{ws} denote two lists that have the same
length, and let \inlinecoq{xs} denote another list.
\begin{itemize}[leftmargin=3.5mm]

\item
  Should convolving \inlinecoq{vs} and \inlinecoq{xs ++ ws} reduce to
  convolving \inlinecoq{vs} and \inlinecoq{ws}, then when reaching the
  end of \inlinecoq{vs}, one should continue to slide through the second
  list in the manner of \sectionref{sec:TAFA}, \ie, in synchrony with
  sliding through \inlinecoq{xs ++ ws} to reach \inlinecoq{ws}, and then
  apply the continuation to \inlinecoq{ws} and the empty list.

\item
  Should convolving \inlinecoq{xs ++ vs} and \inlinecoq{ws} reduce to
  convolving \inlinecoq{vs} and \inlinecoq{ws}, then the continuation
  should test whether its first argument is empty and keep going with the
  second.

\end{itemize}

\noindent
And indeed,
convolving \inlineml{[2; 3; 4; 5]} and \inlineml{[40; 50]}
and
convolving \inlineml{[4; 5]} and \inlineml{[20; 30; 40; 50]}
yield the same result as
convolving \inlineml{[4; 5]} and \inlineml{[40; 50]} in this case:
the prefix of the longer list is ignored, the convolving function is structurally
recursive, and true to TABA
the convolving function traverses the first list at call time
and the second at return time.

\section{Conclusion}
\label{sec:conclusion}

\begin{flushright}
``...if you're in any way excited \\
by the weird and wonderful algorithms \\
we use in functional languages \\
to do simple things like reversing a list.''~\cite{Kidney:20_typing-taba}
\end{flushright}

\noindent
What have we learned here?
\begin{itemize}[leftmargin=3.5mm]

\item
  that the TABA recursion pattern can be further illustrated, that its
  telling example can be refined, and that this refinement suggests
  alternative solutions;

\item
  that TABA can be formalized in a way that crystallizes both its control
  flow and its data flow; and

\item
  that TABA lends itself to an iterative variant, TAFA.

\end{itemize}

\noindent
Proving TABA programs in direct style is carried out equationally and
by structural induction.
Proving TABA programs that use continuations often requires relational
reasoning as well, to characterize these continuations, but
equational reasoning turned out to be sufficient here.

To close, let us turn to the issue of efficiency.
Is it more efficient to use TABA or to construct an intermediate data
structure?
The inter-derivation described in \sectionref{sec:convolving-lists-when-they-are-supposed-to-have-the-same-length}
shows that modulo any representational change of the defunctionalized
continuation (\eg, going from Peano numbers to binary integers), the
time complexity is the same either way.
The answer therefore depends on the underlying implementation of the
language processor such as unboxing polymorphic values in activation
records~\cite{PeytonJones:PR02}.
Moving from quantitative issues to qualitative issues, TABA is noted to
offer an unexpected expressive power in constrained
situations~\cite{Brunel-al:POPL20, Nguyen:PhD, Wang-al:NIPS18}.
Perhaps most significantly, however, TABA sharpens one's understanding of
recursive programming, which is A Good Thing since as is often said,
the sky is the limit once recursion is understood.

\subsection*{Acknowledgments}
Heartfelt thanks to Mayer Goldberg for the original continuation-based implementation of symbolic convolutions
and for our subsequent joint study of its recursion pattern.
The author is also grateful to the anonymous reviewers for perceptive evaluations
and suggestions,
to Bartek Klin and Damian Niminski for their editorship,
and to Julia Lawall for priceless and multi-faceted comments on
two versions
of this article as well as for her playful formalization of \sectionref{sec:reverse2} in
Why3, in the course of the summer of 2020.

\bibliographystyle{fundam}
\providecommand{\noopsort}[1]{}

\appendix

\section{Defunctionalization and refunctionalization}
\label{app:defunctionalization-and-refunctionalization}

Higher-order programs are programs that use functions as values, \eg,
because they involve generic functions such as map or fold or because
they are in continuation-passing style.
First-order programs are programs where values are first-order, \ie, are
not functions.
In the early 1970s~\cite{Reynolds:72}, Reynolds proposed to
`defunctionalize' a particular higher-order program (an interpreter in
continuation-passing style) into a first-order program (a big-step
abstract machine) by representing the continuation with a data type
together with an apply function.
The key idea is that a functional value is an instance of a function
abstraction.
If all the function abstractions that give rise to a functional value can
be enumerated, then the function space is actually a sum type, where each
summand is one of these function abstractions.
Each of these summands can be represented as a
closure~\cite{Landin:CJ64}, \ie, a pair containing the code of the
function abstraction and its lexical environment.
Introducing a functional value (\ie, evaluating a function abstraction)
therefore consists in constructing such a pair, and eliminating it (\ie,
applying a functional value) therefore consists in extending its lexical
environment with the actual parameter(s) and running its code in this
extended environment.
In practice, the code component is replaced by a tag that uniquely
identifies it, and the apply function dispatches on this tag.
Also, this tag is represented as a data-type constructor.

\subsection{An example of defunctionalization}
\label{subapp:an-example-of-defunctionalization}

Consider the continuation-passing implementation from
\sectionref{subsec:refunctionalizing-the-lightweight-fused-implementation-of-rev2}:
\inputcoqrevtwo{REVTWOPRIME_VTWO}
\vspace{-1mm}
\inputcoqrevtwo{REVTWO_VTWO}

\noindent
The continuation has type \inlinecoq{list V -> bool}.
This type is inhabited by instances of two function abstractions:
\begin{itemize}[leftmargin=3.5mm]

\item
  one in \inlinecoq{rev2_v3}: %
  \inlinecoqfootnotesize{fun ws => match ws with | nil => true | w :: ws' => ... end}, and

\item
  one in \inlinecoq{rev2'_v3}: %
  \inlinecoqfootnotesize{fun ws => match ws with | nil => false | w :: ws' => ... end}.

\end{itemize}

\noindent
The resulting functional values are applied in the nil case and in the cons case of \inlinecoq{rev2'_v3}.

\medskip
Since continuations are instances of two function abstractions,
their type can be represented using a data type with two constructors:
\inputcoqrevtwo{CONT_TEMPLATE}

The first function abstraction (in \inlinecoq{rev2_v3}) has no free
variables and the second one (in \inlinecoq{rev2'_v3}) has two,
\inlinecoq{v} and \inlinecoq{h_vs_op}.
Therefore the first constructor has no argument and the second has two:
\inputcoqrevtwo{CONT}

\noindent
The corresponding dispatch function and defunctionalized program read as follows:
\inputcoqrevtwo{DISPATCH_CONT}
\vspace{-1mm}
\inputcoqrevtwo{REVTWOPRIME_VTWO_DEF}
\vspace{-1mm}
\inputcoqrevtwo{REVTWO_VTWO_DEF}

%

\noindent
The alert reader will have noticed that since \inlinecoq{dispatch_cont}
returns a function, the defunctionalized program is still higher-order.
However, its two invocations are fully applied,
so the (higher-order) dispatch function is better defined as a (first-order) apply function:
\inputcoqrevtwo{APPLY_CONT}
\vspace{-1mm}
\inputcoqrevtwo{REVTWOPRIME_VTWO_DEFPRIME}
\vspace{-1mm}
\inputcoqrevtwo{REVTWO_VTWO_DEFPRIME}

\noindent
And since the type \inlinecoq{cont} is isomorphic to that of lists,
that is how the defunctionalized continuation is represented,
which leads one to the implementation of
\sectionref{subsec:lightweight-fusing-the-implementation-in-two-passes-of-rev2}
where \inlinecoq{apply_cont} is named \inlinecoq{rev2''_v2}.

\vspace{-1mm}

\subsection{Refunctionalization}
\label{subapp:refunctionalization}

Refunctionalization is a left inverse of defunctionalization~\cite{Danvy-Millikin:SCP09-short}.
When a data structure is constructed once and dispatched upon once,
chances are it is the first-order counterpart of a higher-order function
and lies in the image of defunctionalization.
For example, the implementation in
\sectionref{subsec:implementation-of-rev2-in-two-passes-that-reverses-the-first-list}
page~\pageref{subsec:implementation-of-rev2-in-two-passes-that-reverses-the-first-list}
is in defunctionalized form, since the sole reason of being for \inlinecoq{vs_op}
is to be processed in \inlinecoq{rev2''_v1}, which de facto serves as its
apply function.
The refunctionalized counterpart reads as follows:
\inputcoqrevtwo{REVTWOPRIME_VZERO_REFUNCT}
\vspace{-1mm}
\inputcoqrevtwo{REVTWO_VZERO_REFUNCT}

\vspace{-1mm}

\subsection{Significance of defunctionalization and refunctionalization}

Reynolds's defunctionalization of a definitional interpreter that had been CPS-transformed
with call by value in mind led to the CEK machine~\cite{Felleisen-Friedman:FDPC3},
and Schmidt's defunctionalization of a definitional interpreter that had been CPS-transformed
with call by name in mind led to the Krivine machine~\cite{Schmidt:HOSC07}
before either of these machines had a name.
Many other abstract machines and first-order algorithms also fit
as well as data structures, be they zippers or evaluation contexts.
In Computer Science, do we invent things or do we discover them?


\section{Generic programming with lists}
\label{app:generic-programming-with-lists}

The functionals \inlinecoq{list_fold_right} and \inlinecoq{list_fold_left}
were originally conceived by Strachey~\cite{Strachey:61}
to program generically over lists: \vspace*{1mm}
\inputcoqrevtwo{LIST_FOLD_RIGHT}
\inputcoqrevtwo{LIST_FOLD_LEFT}

\noindent\vspace*{1mm}
The first captures the ordinary recursive descent over a list
(\eg, instantiating it with nil and cons yields the list-copy function)
and the second the ordinary tail-recursive descent over a list using an accumulator
(\eg, instantiating it with nil and cons yields the list-reverse function).
Each can simulate the other by threading an accumulator.

These two functionals come handy, \eg, as a litmus test for a beginning functional programmer
to demonstrate that their implementation is structurally recursive: if they cannot express it
using a fold function, it isn't.

\medskip
In the present case, all the structurally recursive implementations are
fold-right ready, starting with the very first self-convolution functions
in \sectionref{sec:scnv}.
Here are their fold-right counterparts:
\inputocamlscnv{SELF_CNV_RIGHT}
\vspace{-1mm}
\inputocamlscnv{SELF_CNV_C_RIGHT}

Likewise,
the implementation in
\appendixref{subapp:refunctionalization} is fold-left ready.
Here is its fold-left counterpart:
\inputcoqrevtwo{REVTWO_VZERO_REFUNCT_LEFT}

\noindent
This
counterpart illustrates the rendering of TABA using
fold-left, a point recently made about convolving lists in
Kidney's scientific blog~\cite{Kidney:20_typing-taba}.

\section{Lambda-lifting and lambda-dropping}
\label{app:lambda-lifting-and-lambda-dropping}

Block structure and lexical scope are two cornerstones of functional programming,
witness the definition of \inlinecoq{list_fold_right} and \inlinecoq{list_fold_left}
in \appendixref{app:generic-programming-with-lists}, where \inlinecoq{visit} is defined locally
in the scope of \inlinecoq{nil_case} and \inlinecoq{cons_case}.
A popular alternative is recursive equations: global mutually recursive functions
with no local declarations.
For example, here is an implementation of \inlinecoq{list_fold_right} as a recursive equation:
\inputcoqrevtwo{LIST_FOLD_RIGHT_LIFTED}

When the world was young~\cite{PeytonJones:87}, recursive equations
were found to be a convenient format for compiling functional programs
for the G-machine.
A program transformation, lambda-lifting~\cite{Johnsson:FPCA85}, was developed
that parameterized each local function with its free variables,
thus making them scope insensitive, which made it possible for them to float up
to the top level and become (mutually) recursive equations.
Compilers then evolved, and it was found beneficial to lambda-drop functional programs
into programs with more block structure and more free variables~\cite{Danvy-Schultz:TCS00-short},
on the ground that they could be both compiled and executed more efficiently.
And indeed compare the lambda-dropped version of \inlinecoq{list_fold_right}
and its lambda-lifted version, \inlinecoq{list_fold_right'}:
in the lambda-dropped version, the auxiliary function has one parameter,
and in the lambda-lifted version, it has many more, most of them unchanging.

Nowadays the issue is moot for programming,
since efficient compilers are known to lambda-drop source programs internally~\cite{Dybvig:ICFP06-short}.

\medskip
For proving, however, lambda-lifted programs are more convenient to reason about
since we can only refer to entities by their name
and we can only mention names that are defined globally.

\section{Tail calls, non-tail calls, lightweight fusion, and lightweight fission}
\label{app:lightweight-fusion-and-lightweight-fission}

Figure~3 displays four implementations to compute the first and last elements of a non-empty list.
If the given list is empty, the result is \inlinecoq{None}.
Otherwise, we already know the first element, and all we need to do is to compute the last element,
which is carried out tail-recursively by an auxiliary function
that is defined by structural induction over the tail of the given list.

\medskip
The two first implementations are lambda-dropped in that their auxiliary function is local
(and \inlinecoq{v} is declared in an outer scope),
and the two last implementations are lambda-lifted in that their auxiliary function is global.
The first and the third implementations are lightweight-fissioned in that the auxiliary function is invoked with a non-tail call,
and the second and the fourth implementations are lightweight-fused in that the auxiliary function is invoked with a tail call.
In the last implementation, \inlinecoq{v} is passed as an extra parameter to the auxiliary function.
The codomains of the auxiliary functions differ depending on whether they are fused or fissioned.

Lightweight fusion by fixed-point promotion is due to Ohori and Sasano~\cite{Ohori-Sasano:POPL07}.
Lightweight fission by fixed-point demotion is its left inverse.

\begin{figure}[!htbp]
\vspace{-5mm}
\inputcoqfal{FIRST_AND_LAST_DROPPED_FISSIONED}
\vspace{-2mm}
\inputcoqfal{FIRST_AND_LAST_DROPPED_FUSED}
\vspace{-2mm}
\inputcoqfal{FIRST_AND_LAST_LIFTED_FISSIONED_AUX}
\vspace{-2mm}
\inputcoqfal{FIRST_AND_LAST_LIFTED_FISSIONED}
\vspace{-2mm}
\inputcoqfal{FIRST_AND_LAST_LIFTED_FUSED_AUX}
\vspace{-2mm}
\inputcoqfal{FIRST_AND_LAST_LIFTED_FUSED}
\vspace{-2mm}
\caption{Four implementations to compute the first and last elements of a non-empty list}
\label{fig:first_and_last}
\end{figure}

 \medskip
The TAFA example in
\sectionref{subsubsec:list-index-rtl-formalizing-and-proving} is a good
candidate for lightweight fusion:
\begin{itemize}[leftmargin=3.5mm]
\item
  since \inlinecoq{list_index_rtl_there} and
  \inlinecoq{list_index_rtl_forth} are tail recursive, we can
  relocate the context of their initial call (\ie, the match expression
  in \inlinecoq{list_index_rtl}) into their body, making the
  definition of \inlinecoq{list_index_rtl} tail-recursive:

\inputcoqindex{LIST_INDEX_RTL_TAFAP}

 \eject  
\item
  in \inlinecoq{list_index_rtl'_there},
  the match expression from \inlinecoq{list_index_rtl}
  is relocated around the two possible end results of this tail-recursive
  function:
\inputcoqindex{LIST_INDEX_RTL_TAFAP_THERE}

\item
  in \inlinecoq{list_index_rtl_forth'},
  the match expression from \inlinecoq{list_index_rtl}
  is relocated around the two possible end results of this tail-recursive
  function:
\inputcoqindex{LIST_INDEX_RTL_TAFAP_FORTH}
\end{itemize}

\vspace{-2mm}

\noindent
Note how the codomain of each functions is now the same (namely that of
the final result), and how \inlinecoq{list_index_rtl_there'}
and \inlinecoq{list_index_rtl_forth'} take extra parameters since
these parameters are now no longer available in the lexical environment,
two byproducts of the program now being tail-recursive.
As for the type annotations (\eg, \inlinecoq{@None(List V)}), they were
added to appease the type inferencer.

The relocated match expressions are then simplified, yielding the
following tail-recursive program:
\inputcoqindex{LIST_INDEX_RTL_TAFAQ_FORTH}
\vspace{-1mm}
\inputcoqindex{LIST_INDEX_RTL_TAFAQ_THERE}
\vspace{-1mm}
\inputcoqindex{LIST_INDEX_RTL_TAFAQ}

\noindent
Likewise, in
\sectionref{subsec:computing-the-common-suffix-of-two-lists}, the
\inlinelisp{slide} procedure is a candidate for lightweight fusion.

\section{A vademecum for continuations}
\label{app:continuations-in-a-nutshell}

\subsection{Direct style vs. Continuation-Passing Style (CPS)}
\label{subapp:direct-style-vs-continuation-passing-style}

An expression is said to be in ``direct style'' if its evaluation is
recursive and its intermediate results are not named.
For example, given appropriately typed \inlineocaml{f}, \inlineocaml{g},
and \inlineocaml{x}, the expression \inlineocaml{f (g x)} is in direct style.
An expression is in ``monadic style'' if its evaluation is tail recursive
and its intermediate results are named.
For example, assuming call by value, both \inlineocaml{let v1 = g x in f v1}
and \inlineocaml{let v1 = g x in let v2 = f v1 in v2} are in monadic
style.
And it is in ``continuation-passing style'' if its evaluation is tail recursive
and if the functions it involves take an extra argument, the continuation.
For example, still assuming call by value, both \inlineocaml{fun k -> g_cps x (fun
  v1 -> f_cps v1 k)} and \inlineocaml{fun k -> g_cps x (fun v1 -> f_cps v1 (fun v2 ->
  k v2))} are in
CPS,
where \inlineocaml{f_cps} and \inlineocaml{g_cps} are the
continuation-passing counterpart of \inlineocaml{f} and \inlineocaml{g}.

\medskip
For any evaluation order, any expression can be CPS-transformed using the
explanation in the previous paragraph: (1) name intermediate results; (2)
sequentialize their continuation by re-associating the resulting
let-expressions to flatten them and $\eta$ reduce the inner one, for the
sake of proper tail recursion; and (3) introduce continuations.
For example, here is the continuation-passing counterpart of \inlineocaml{list_index_rtl_ds}
in \sectionref{subsubsec:list-index-rtl-programming}:
\inputocamlindex{LIST_INDEX_RTL_CPS}

\noindent
The corresponding trace is a tail-recursive counterpart of that in
\sectionref{subsubsec:list-index-rtl-programming}, namely a complete
series of tail calls to \inlineocaml{visit} where continuations are
accumulated, followed by a complete series of tail calls to these
accumulated continuations when the list is too short, and a complete
series of tail-calls to \inlineocaml{visit} where continuations are
accumulated, followed by a complete series of tail calls to these
accumulated continuations when the list is long enough:
\inputocamlindex{TRACED_LIST_INDEX_RTL_CPS}

\subsection{Continuations and their scope}

Consider the identifiers that name continuations in a continuation-passing expression.
In the image of the CPS transformation, one identifier is enough:
continuations are declared and then used linearly in a LIFO manner~\cite{Danvy:SCP94}.
(Then defunctionalizing a continuation gives rise to a stack.
And when the program in CPS is an evaluator,
this stack is known as ``the control stack'' since Dijkstra~\cite{Dijkstra:60}.)

\medskip
Sometimes, though, it is not the current continuation that is applied to
continue the computation, but another one that was declared elsewhere,
\eg, earlier~\cite{Danvy-Lawall:LFP92}.
Applying this other continuation discontinues the current computation and
makes it continue elsewhere or earlier.
Or sometimes, the continuation is not applied at all, indicating that
it is delimited (\ie, an initial continuation is provided somewhere in the program).
Not applying any continuation discontinues the current computation and makes it stop
and return an intermediate result to the point where the initial continuation was provided.
The control effect that is being emulated there
is that of an exception in direct style (see next section for a concrete example).
Many other control effects can be emulated using continuations; they give rise
to control operators in direct style that achieve the corresponding control effect,
\eg, delimited control~\cite{Danvy-Filinski:LFP90} as well as
computational monads~\cite{Filinski:POPL94}.

\subsection{Continuation-passing vs.\ continuation-based programs}

In a nutshell, the co-domain of a continuation-passing function is
polymorphic and the co-domain of a continuation-based function is not.

\medskip
Consider, for example, the traditional factorial function in
continuation-passing style.
Its continuation is linear and used in a LIFO manner:
\inputocamlcpsvscb{FAC_CPS}

\noindent
This tail-recursive implementation is interfaced with the direct-style
world by supplying it with an initial continuation that is the identity
function, which delimits the continuation and instantiates the type
variable to \inlineocaml{int}:
\inputocamlcpsvscb{FAC}

Consider, for example, a tail-recursive implementation of a function that
detects whether a binary tree of natural numbers with weightless nodes is
a Calder mobile~\cite{Danvy:RS-04-41}.
Its continuation is affine and used in a LIFO manner:
\inputocamlcpsvscb{BT}
\vspace{-1mm}
\inputocamlcpsvscb{BALANCEDP_CB}
\vspace{-1mm}
\inputocamlcpsvscb{BALANCEDP}

\noindent
This implementation is continuation-based: it is tail recursive and uses
a continuation, but it only uses this continuation as long as the
subtrees that are traversed so far are balanced.
Otherwise, it stops, which commits the co-domain to be
\inlineocaml{bool}.

A continuation-passing program can only be tail recursive -- its
continuations are undelimited.
A continuation-based program, on the other hand, need not be tail
recursive -- its continuation is delimited and therefore can be composed.

\medskip
Consider, for example, a function that maps a list of unknown length $n$
to the list of its prefixes.
Its continuation is delimited and non-linear:
\inputocamlcpsvscb{TEST_PREFIXES}
\vspace{-1mm}
\inputocamlcpsvscb{PREFIXES_CB}
\vspace{-1mm}
\inputocamlcpsvscb{PREFIXES}

\noindent
Ostensibly, it proceeds in $n$ recursive calls, even though the size of
its result is quadratic in $n$:
\inputocamlcpsvscb{PREFIXES_WITH_FOLD}

\noindent
The key is to compose each of its successive continuations to construct
the successive prefixes, something that can also be achieved in CPS by
layering continuations~\cite{Danvy-Filinski:LFP90}:
\inputocamlcpsvscb{PREFIXES_CPS}
\vspace{-1mm}
\inputocamlcpsvscb{PREFIXESP}

\vspace{-1mm}

\subsection{Splitting continuations}
\label{subapp:splitting-continuations}

Based on the type isomorphism between $A + B \rightarrow C$ and $(A
\rightarrow C) \times (B \rightarrow C)$, we can split the continuation
into two in the definition of \inlineocaml{list_index_rtl_cps}
from \appendixref{subapp:direct-style-vs-continuation-passing-style}:
\inputocamlindex{LIST_INDEX_RTL_CPS2}

\noindent
where we can also $\eta$ reduce \inlineocaml{fun v -> k_Found_it v} into \inlineocaml{k_Found_it}.
Instead of passively threading \inlineocaml{k_Found_it} from its point of definition
to its point of use, we can drop this parameter from its point of definition
to its point of use and simplify its application~\cite{Danvy-Schultz:TCS00-short}:
%
%
\inputocamlindex{LIST_INDEX_RTL_CB}

\noindent\smallskip
In this parameter-dropped definition, \inlineocaml{visit} only uses its
continuation until \inlineocaml{n} denotes \inlineocaml{0}, witness the
following trace that features

\begin{itemize}[leftmargin=3.5mm]

\item
a complete series of tail calls to
\inlineocaml{visit} where continuations are accumulated, followed by a
complete series of tail calls to these accumulated continuations when the
list is too short, and

\item
a complete series of tail-calls to
\inlineocaml{visit} where continuations are accumulated, followed by an
interrupted series of tail calls to these accumulated continuations when
the list is long enough:
\end{itemize}

\inputocamlindex{TRACED_LIST_INDEX_RTL_CB}

\noindent\vspace{1mm}
Discontinuing the computation is characteristic of encoding an exception
in direct style.
Exceptions in OCaml, however, are global and monomorphic, so to
preserve the polymorphism of the corresponding direct-style
implementation, one needs to resort to a parameterless exception and a
local \linebreak reference:

\eject
\inputocamlindex{FOUND_IT}
\vspace{-1mm}
\inputocamlindex{LIST_INDEX_RTL_DSE}

\medskip\noindent
The traces are as expected, namely
\begin{itemize}[leftmargin=3.5mm]

\item
a complete series of recursive calls
to \inlineocaml{visit} followed by a complete series of returns when the
list is too short, and

\item
a complete series of recursive calls to
\inlineocaml{visit} followed by an interrupted series of returns when the
list is long enough:
\end{itemize}

\inputocamlindex{TRACED_LIST_INDEX_RTL_DSE}

\clearpage

\begin{frameit}
\vspace{-4mm}
\begin{quote}
Dear Reader:

Thanks for getting acquainted further with ``There and Back Again.''
As a farewell gift, here are some more programming exercises.

\begin{description}[leftmargin=3.5mm]
\itemsep=0.9pt
\item[Reversing a list, again:] \ \\
Given a list of length $n$, where $n$ is unknown, construct its reverse in
$n$ recursive calls without using an accumulator.
Not using an accumulator means that when expressing your solution using
either fold functional for lists (\ie, \inlineocaml{list_fold_right} or
\inlineocaml{list_fold_left}, they give the same result here), the first
argument of the fold functional should not be a function.

\item[Deciding whether a list is a self-convolution:] \ \\
Given a list of pairs of length $n$, where $n$ is unknown,
determine whether this list represents a self-convolution in $n$
recursive calls.

\item[Decomposing a symbolic convolution into its two components:] \ \\
Given a symbolic convolution $[(x_1, y_n), (x_2, y_{n-1}), ..., (x_{n-1},
  y_2), (x_n, y_1)]$, where $n$ is unknown, construct its two components
$[x_1, x_2, ..., x_n]$ and $[y_1, y_2, ..., y_n]$ in
$n$ recursive calls.

\item[Swapping parity-indexed elements in lists of odd length:] \ \\
Implement a function that, given a list of odd length, swaps its
even-indexed elements, so that, e.g.,
\inlineocaml{[0; 1; 2; 3; 4; 5; 6]}
is mapped to \inlineocaml{Some [6; 1; 4; 3; 2; 5; 0]}, and another that,
given a list of odd length, swaps its odd-indexed elements, so that,
e.g., \inlineocaml{[0; 1; 2; 3; 4; 5; 6]} is mapped to \inlineocaml{Some
  [0; 5; 2; 3; 4; 1; 6]}.
Given a list of length $n$, where $n$ is unknown, the two functions should
proceed in $n$ recursive calls.
Any list of even length should be mapped to \inlineocaml{None}.

\item[Swapping parity-indexed elements in lists of even length:] \ \\
Guess what.

\item[Twice as fast:] \ \\
Assume a function \inlineocaml{rev_stutter2} that maps a list
  $[x_1, x_2, ..., x_{n-1}, x_n]$ to
  $[x_n, x_n,$ $x_{n-1}, x_{n-1}, ..., x_2, x_2, x_1, x_1]$.
Given two lists, the first one of length $n$, where $n$ is unknown,
detect whether the second list is the result of applying
\inlineocaml{rev_stutter2} to the first list, in $n$ recursive calls.

At least two solutions exist: one that recurses on the first list, and
another that recurses on the second list.
\end{description}
\noindent
Do feel free to share your solutions with the author by email, just
for the joy of functional programming and proving.
There are no open problems here: the point of these exercises is that if
you can solve any of them, you do get TABA.
\end{quote}
\vspace{2.5mm}
\end{frameit}

\end{document}